\documentclass[preprint]{revtex4}
\usepackage{graphicx}
\usepackage{rotating}
\usepackage{color}
\begin{document}
\title{Spins in Exotic Nuclei:\\ RI-beam Experiments with Polarized Targets}
\author{T.~Uesaka}
\affiliation{RIKEN Nishina Center for Accelerator-Based Science, Saitama 351-0198, Japan\\ \email{uesaka@riken.jp}}

\date{Received: date/ Revised version: date}
\begin{abstract}
 Spin-degrees of freedom play a significant role in exotic nuclei. Scattering with polarized protons has potential as a powerful tool to explore the spin effects in nuclei. This lecture note discusses the background, current status, and future prospects of experimental studies with spin polarized targets and radioactive ion beams.
\end{abstract}

\maketitle

\section{Spin degrees of freedom in exotic nuclei}

Since introduction of radioactive ion (RI) beams 30 years ago, nuclear physicists have witnessed the significant role of spin-degrees of freedom in exotic nuclei. Today it is widely recognized that the effects of spin-dependent interactions, such as spin-orbit, tensor, and three-nucleon interactions, are essential to understand the structure and reactions of exotic nuclei. Although the spin-dependent effects can be investigated through experimental data of masses, level schemes, electromagnetic moments etc., more direct evidence of the effects should be acquired through spin observables in the reactions of exotic nuclei. In this lecture note, the background, current status, and future prospects of experimental studies with spin polarized targets and RI beams are discussed.

\section{Spin effects in nuclei}
\subsection{Spin-orbit coupling and magic numbers in nuclei \label{sec:spinorbithistory}}
Various many-Fermion systems form a ``shell'' structure, including some atoms, atomic clusters, quantum dots, and nuclei. Due to the shell formation, such systems become stable when constituent particles have certain particular numbers, known as ``{\it magic numbers}''.  What is interesting is that each system has different magic numbers: 2, 10, 18, 36, 54, 86, 118 . . . for atoms, 2, 8, 20, 40, 58, 92, 138 for atomic clusters~\cite{Brack93}, and 2, 8, 20, 28, 50, 82, 126 . . . for nuclei. Thus the magic numbers act as fingerprints of many-Fermion systems and reveal the nature of the system.

\begin{figure}[htbp]
  \centering
  \resizebox{0.6\textwidth}{!}{\includegraphics{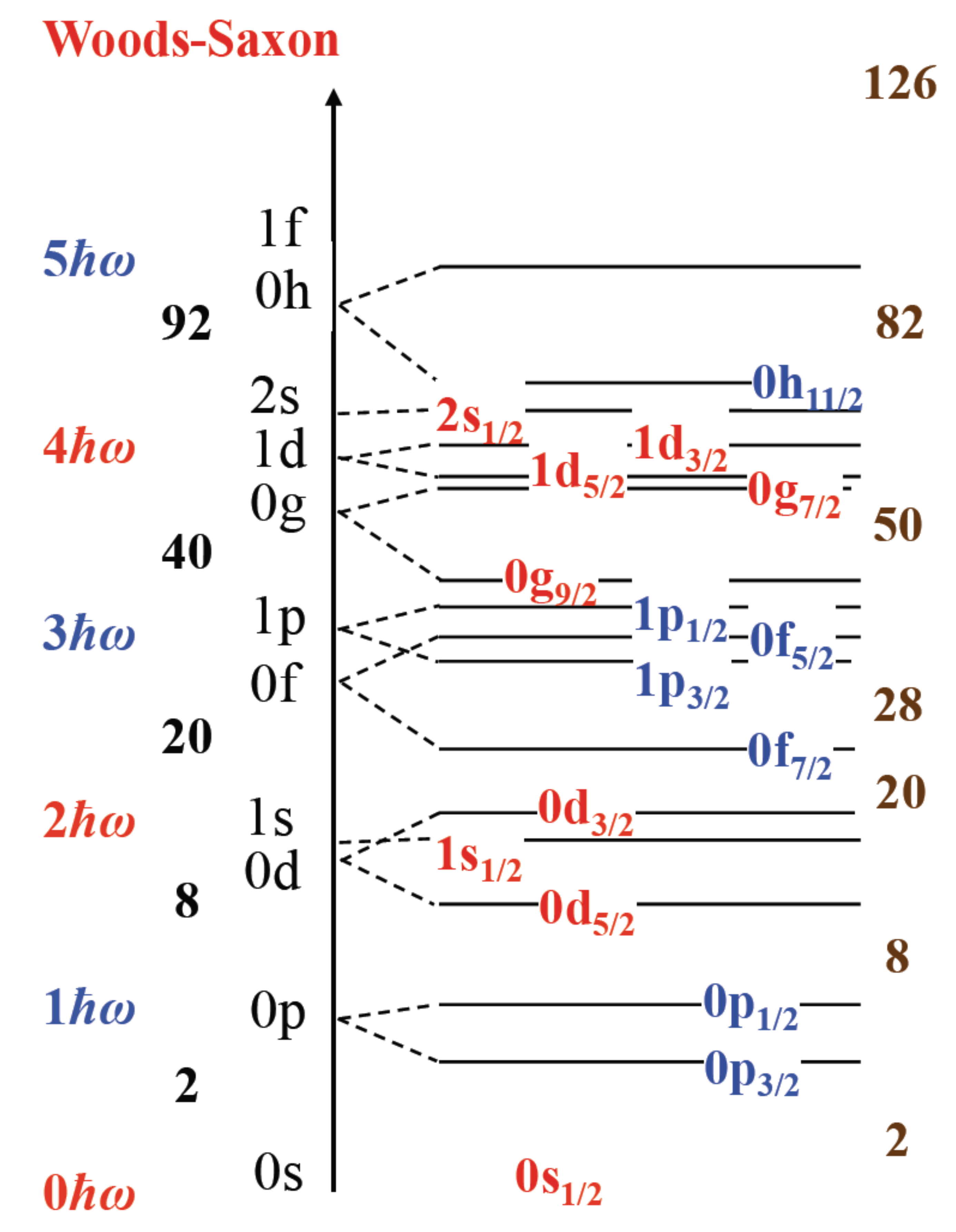}}
  \caption{(color online) Nuclear shell structure and magic numbers. \label{fig:shellstructure}}
\end{figure}

The formation mechanism of the nuclear magic numbers was initially a mystery. 
Many nuclear theorists tried to explain the magic numbers in terms of a simple-minded potential
model, but all the efforts failed until the end of 1940's. In 1949, Mayer~\cite{Mayer49} and Haxel, Jensen and Suess~\cite{Haxel49} proposed the introduction of a strong spin-orbit coupling to solve the mystery and successfully explained the magic numbers 28, 50, 82, and 126. As shown in Fig.~\ref{fig:shellstructure}, lowering of the $j=\ell+1/2$ orbits with large $\ell$ ($f_{7/2}$, $g_{9/2}$, $h_{11/2}$) caused by a strong spin-orbit coupling results in the shell closure at 28, 50, and 82.

Although calculations with the strong spin-orbit coupling explained the observations well, it was unclear whether the nuclear interaction has the strong spin-orbit term required by the shell formation. The situation can be seen in Ref.~\cite{Mayer50} where  Mayer claimed ``{\it There is no adequate theoretical reason for the large observed value of the spin orbit coupling. The Thomas splitting has the right sign, but is utterly inadequate in magnitude to account for the observed values.}''\footnote{Early history of the nuclear spin-orbit splitting can be found in Ref.~\cite{BarschallBrown86}.}

\begin{figure}[htbp]
  \centering
  \resizebox{0.6\textwidth}{!}{\includegraphics{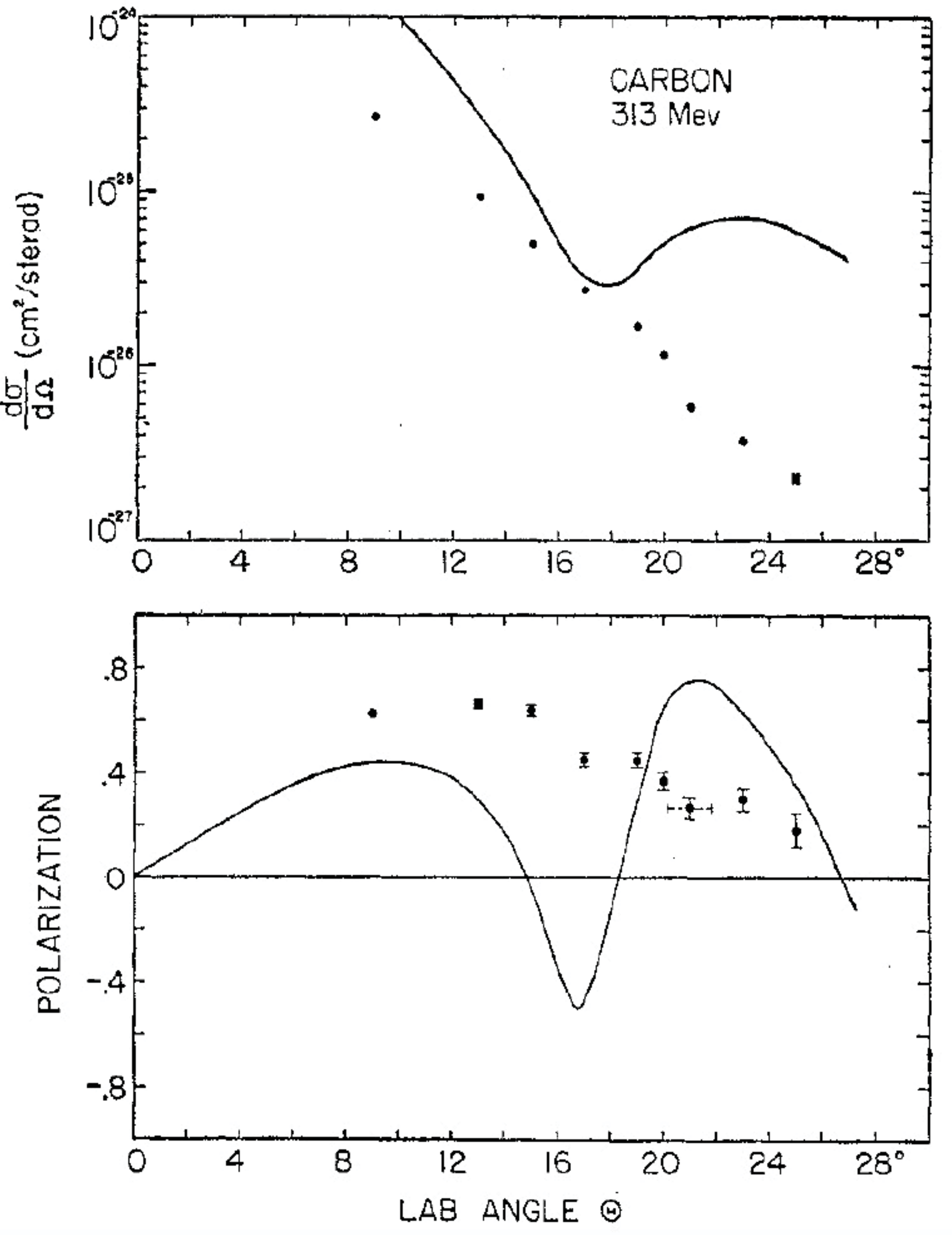}}
  \caption{Cross section (top) and polarization (bottom) in the $p$-${\rm ^{12}C}$ elastic scattering at $E_{p}=313$~MeV. Experimental data are shown with points. Curves are results of a phase-shift calculation for a model proposed by Fermi. Reprinted from Ref.~\cite{Chamberlain56}\copyright 1956 with permission from the American Physical Society. \label{fig:Chamberlain}}
\end{figure}

This situation motivated experimentalists to measure spin observables in proton-nucleus scatterings which give access to the magnitude of nuclear spin-orbit interactions. Since a spin-polarized beam did not exist at that time, a ``double scattering'' method was employed to measure the spin observables~\cite{Chamberlain56,Heusinkveld52,Oxley53,Chamberlain57}. 
Figure~\ref{fig:Chamberlain} shows the induced polarization in the proton-${\rm ^{12}C}$ elastic scattering at a proton incident energy of 313~MeV~\cite{Chamberlain56}. The obtained polarization has a value as large as $\sim$0.7, which is a clear manifestation of the strong spin-orbit interaction.

Fermi analyzed the data by means of simple perturbation
calculations and concluded that the strength of spin-orbit coupling assumed in the
shell model is consistent with that required to produce the polarization in proton
scattering~\cite{Fermi54}. At this point, the existence of the strong spin-orbit coupling in nuclei was established. However, the microscopic origin of the nuclear spin-orbit potential was still not clear.

\subsection{Microscopic origins of the strong spin-orbit coupling}

 Prior to the proposals by Mayer and Jensen, Inglis investigated the microscopic origins of nuclear spin-orbit coupling using an analogy to the atomic spin-orbit coupling~\cite{Inglis36}. Inglis considered that the relativistic effect, which is usually referred to as the Thomas effect~\cite{Thomas26}, should be dominant over the magnetic contribution in nuclear systems and introduced the concept of ``an inversion doublet'', namely an orbit with $j=\ell+1/2$ locates lower in energy than that with $j=\ell-1/2$. The order of $j=\ell+1/2$ and $j=\ell-1/2$ orbits is the opposite of that in atoms and is consistent with observations. Inglis evaluated the interval between the spin doublet, which is known as {\it spin-orbit splitting}, only with the Thomas effect and found that the calculation agreed with experimental results for ${\rm ^{7}Li}$. However, it was found later that this agreement was accidental and the spin-orbit splitting due to the Thomas effect is too small in most nuclei. Today, the small spin-orbit splitting in ${\rm ^{7}Li}$ is considered to be due to its established cluster structure.
  
 The first theoretical model of nuclear spin-orbit coupling was proposed by Fujita and Miyazawa in 1957~\cite{FujitaMiyazawa57-2}, immediately after their famous work on a three-nuclear force via an intermediate $\Delta$ excitation~\cite{FujitaMiyazawa57}. They found that the three-nucleon force with an intermediate $\Delta$ excitation results in a spin-orbit coupling and concluded that `` {\it our spin-orbit coupling derived from the meson-theoretical three-body force is about 4.3 times larger than the Thomas term. The existence of the correlation may also increase this magnitude further. It is probable that the higher-order corrections for three-body forces have a large effect}''. 
 
  \begin{figure}[htbp]
  \centering
  \resizebox{0.8\textwidth}{!}{\includegraphics{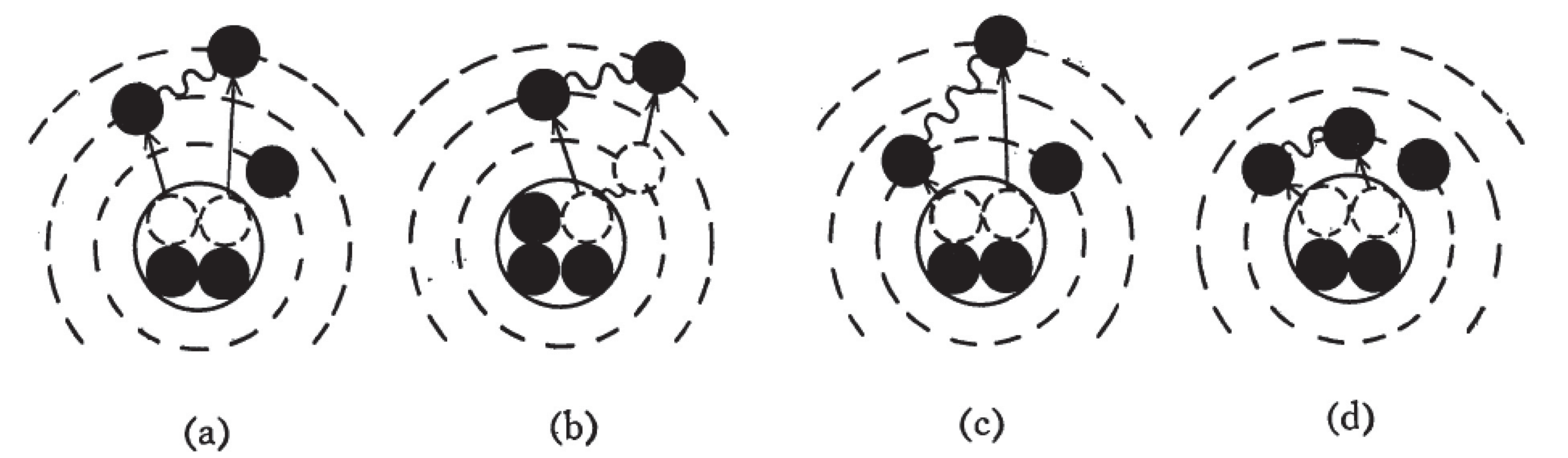}}
  \caption{Configurations of proton-neutron pair excitations: (a) both proton and neutron from the 0s-shell, (b) one from the 0s and the other from 0p. (c) and (d) similar to (a), but one or two nucleons suffer from the Pauli-blocking in the 0p-shell.  Reprinted from Ref.~\cite{Terasawa60}\copyright 1960 with permission from the Oxford University Press. \label{fig:Terasawa}}
\end{figure}  

Secondly, the tensor-force contribution on the spin-orbit coupling was analyzed by Terasawa~\cite{Terasawa60}. The tensor force virtually excites a proton-neutron pair to high-lying levels and the resulting second-order perturbation effects account for the major part of the nuclear binding energy. This mechanism, which is the most important in nuclear physics, will be explained more in detail in the next subsection. Terasawa investigated the cases of ${\rm ^{5}He}$ and ${\rm ^{15}N}$. Figure~\ref{fig:Terasawa} depicts the former where the smallest circle at the center indicates the 0s-shell and the outer circles are higher shells. In Fig.~\ref{fig:Terasawa}(a), the proton-neutron pair in the 0s-shell is virtually excited to the high-lying shells, while in Fig.~\ref{fig:Terasawa}(b) one nucleon in the 0s-shell and another in the 0p-shell are virtually excited. The cases in Fig.~\ref{fig:Terasawa}(c)  and (d) are similar to that in Fig.~\ref{fig:Terasawa}(a), but the Pauli blocking effect in the 0p-shell is considered. Terasawa showed that the contributions of Fig.~\ref{fig:Terasawa}(c)  and (d) are larger than those from (a) and (b), resulting in a larger attraction in the state with $j=\ell +1/2$. The calculated spin-orbit splitting is about a half of the observed value.

Later, meson exchange theories of nuclear forces elucidated that two-nucleon spin-orbit interaction arises from $\sigma$ and $\omega$ meson exchanges~\cite{Machleidt89}.  According to the Bruckner-Hartree-Fock calculation in Ref.~\cite{Scheerbaum76}, the two-body spin-orbit interaction explains roughly half of the observed spin-orbit splittings. 

The contributions of the three-nucleon, tensor, and the two-nucleon spin-orbit interactions in medium heavy nuclei 
were investigated by Ando and Bando~\cite{AndoBando81} and Pieper and Pandharipande~\cite{Pieper93}.
They have calculated the spin-orbit splitting in double magic ${\rm ^{16}O}$ and ${\rm ^{40}Ca}$ nuclei and concluded that the contributions from the three-nucleon and tensor interactions are 20--30\% each, while that from the two-nucleon spin-orbit interaction is approximately half of the total splitting. 

\subsection{Tensor-force effects in nuclei}
The previous section introduced the tensor-force contribution to the nuclear spin-orbit coupling. Since the tensor force is the most important part of the nuclear forces, the effects in the nuclear structure are discussed more in detail in this section.

The story of nuclear tensor forces began with the discovery of a finite quadrupole moment of deuteron by the group lead by Rabi~\cite{Kellogg39}. Immediately after the report, Schwinger elucidated that the finite quadrupole moment means the existence of a tensor force with an operator $S_{12}=3(\sigma_{1}\cdot r)(\sigma_{2}\cdot r)/r^2-\sigma_{1}\cdot\sigma_{2}$~\cite{Schwinger39}.
Later a pion-exchange model of the nuclear force by Yukawa~\cite{Yukawa35} clearly demonstrated that a strong tensor force is driven by a pion exchange.

To understand how the tensor force contributes to nuclear binding, consider the case of deuteron. The Schr\"odinger equation for a deuteron wave function $\varphi(r)$ can be written as
\begin{eqnarray}
  \left[ -\frac{\hbar^2}{M}\frac{1}{r}\frac{d^2}{dr^{2}} r + \frac{\hbar^2}{M}\frac{\ell(\ell+1)}{r^{2}} +V_{c}+V_{\ell s} \ell\cdot s +V_{t}S_{12}\right] \varphi(r) 
       =\varepsilon_{d}\varphi(r) , 
\end{eqnarray}
%\begin{eqnarray}
%  \left[ -\frac{\hbar^2}{M}\frac{1}{r}\frac{d^2}{dr^{2}} r \right.  &-& \frac{\hbar^2}{M}\frac{\ell(\ell+1)}{r^{2}} \hfill \nonumber \\
%  & &  \hspace{-24pt} +\left. \phantom{\frac{d^2}{dr^{2}}}\hspace{-0.7cm}V_{c}+V_{\ell s} \ell\cdot s +V_{t}S_{12}\right] \varphi(r) 
%       =\varepsilon_{d}\varphi(r) , 
%\end{eqnarray}
where $V_{c}$,  $V_{\ell s}$, and $V_{t}$ are central, spin-orbit, and tensor parts of the nuclear interactions. 
The deuteron wave function is decomposed into $S$- and $D$-state wave function as
\begin{eqnarray}
  r\varphi(r) = u(r){\cal Y}_{0} + w(r){\cal Y}_{2} , 
\end{eqnarray}
where $u(r)$ and $w(r)$ are the $S$- and $D$-state radial wave functions. The angular wave functions ${\cal Y}_{0}$ and ${\cal Y}_{2}$ for the $S$- and $D$-states do not appear in the following discussion.
With the $S$- and $D$-state radial wave functions, the Schr\"odinger equation can be rewritten as 
\begin{eqnarray}
 && \left[ \frac{\hbar^2}{M}\frac{d^2}{dr^{2}}  - \underbrace{V_{c}}_{V_{S}(r)}+\underbrace{\sqrt{8}V_{t}\frac{w(r)}{u(r)}}_{V'_{S}(r)}+\varepsilon_{d}\right] u(r) = 0 , \label{eq:Schrodinger-S}\\
&&  \left[ \frac{\hbar^2}{M}\frac{d^2}{dr^{2}}  - \underbrace{\left(\frac{\hbar^2}{M}\frac{6}{r^2}+V_{c}-3V_{\ell s}-2V_{t}\right)}_{V_{D}(r)}+\underbrace{\sqrt{8}V_{t}\frac{u(r)}{w(r)}}_{V'_{D}(r)} 
+\varepsilon_{d}\right] w(r) = 0 . \label{eq:Schrodinger-D}
\end{eqnarray}
%\begin{eqnarray}
% && \left[ \frac{\hbar^2}{M}\frac{d^2}{dr^{2}}  - \underbrace{V_{c}}_{V_{S}(r)}+\underbrace{\sqrt{8}V_{t}\frac{w(r)}{u(r)}}_{V'_{S}(r)}+\varepsilon_{d}\right] u(r) = 0 , \label{eq:Schrodinger-S}\\
%&&  \left[ \frac{\hbar^2}{M}\frac{d^2}{dr^{2}}  - \underbrace{\left(\frac{\hbar^2}{M}\frac{6}{r^2}+V_{c}-3V_{\ell s}-2V_{t}\right)}_{V_{D}(r)}\right. \nonumber  \\
%&& \left.\hspace{2.5cm} +\underbrace{\sqrt{8}V_{t}\frac{u(r)}{w(r)}}_{V'_{D}(r)} 
%+\varepsilon_{d}\right] w(r) = 0 . \label{eq:Schrodinger-D}
%\end{eqnarray}

In Eqs.(\ref{eq:Schrodinger-S}) and (\ref{eq:Schrodinger-D}), the mixed terms originating from the $S_{12}$ operator are represented in the forms of the pseudo potentials, $V'_{S}(r)$ and $V'_{D}(r)$, with the ratios of $u(r)$ and $w(r)$.  

\begin{figure}[htbp]
 \centering
 \resizebox{0.8\textwidth}{!}{\includegraphics{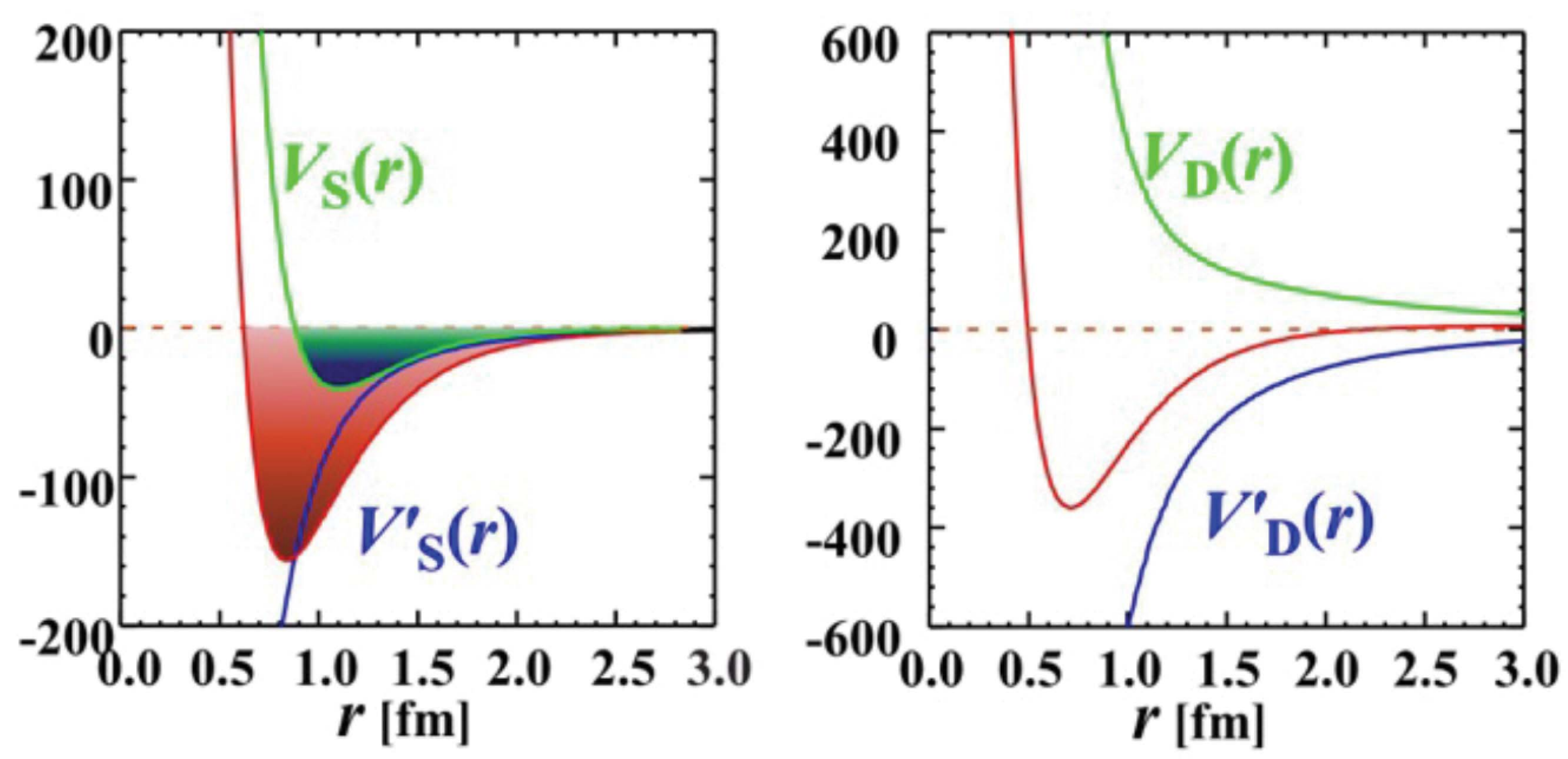}}
 \caption{(color online) Potentials (in MeV) in the $S$- and $D$-states of deuteron. Green, blue, and red curves denote the ordinary potentials [$V_{S}(r)$ and $V_{D}(r)$], the pseudo potentials [$V'_{S}(r)$ and $V'_{D}(r)$] due to the mixing of $S$- and $D$-states, and their sums, respectively. \label{fig:deuteronpotential}}
\end{figure}

Figure~\ref{fig:deuteronpotential} shows the calculations with a realistic nucleon-nucleon interaction. Homogeneous terms, $V_{S}(r)$ and $V_{D}(r)$ (green curves) are slightly attractive for the $S$-state and repulsive for the $D$-state. The attraction in the $S$-state is insufficient to bind the deuteron. On the other hand, the pseudo potentials (blue curves) originating from inhomogeneous terms, $V'_{S}(r)$ and $V'_{D}(r)$, are strongly attractive in both states. Thus the potentials obtained by summing the homogeneous and inhomogeneous terms (red curves) are sufficiently attractive to bind deuteron. Consequently, one can conclude that the inhomogeneous term, which represents a mixing of $S$- and $D$-states driven by the tensor force, is the primary origin of binding in deuteron. This contribution is as much as 70\% of the total attraction in deuteron. 

\begin{figure}[htbp]
 \centering
 \resizebox{0.7\textwidth}{!}{\includegraphics{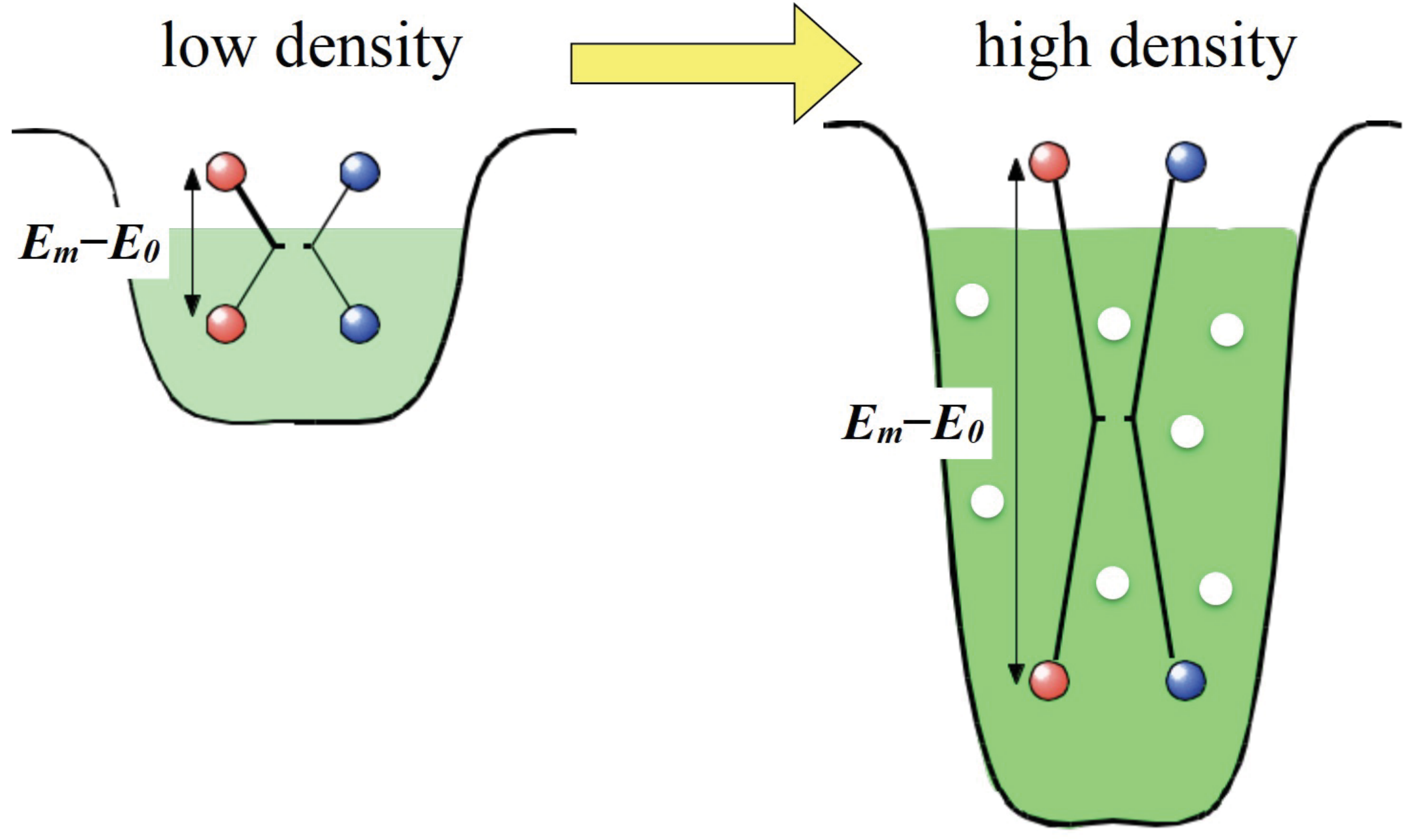}}
 \caption{(color online) Saturation mechanism due to the weakening of the tensor attraction. \label{fig:saturation}}
\end{figure}

This mechanism is responsible for approximately half of the binding energy of other stable nuclei. Below, it is shown that the nuclear saturation at $\rho\sim 0.17$~fm$^{-3}$ can be explained in light of the mechanism.  The energy gain due to the mixing with highly excited states driven by the tensor force can be evaluated by the second order perturbation as 
\begin{eqnarray}
  \Delta E = \sum_{m} \frac{\langle ^{3}S_{1}|V_{t}S_{12}|^{3}D_{1;m}\rangle {\cal Q}\langle ^{3}D_{1;m}|V_{t}S_{12}|^{3}S_{1}\rangle}{E_{m}-E_{0}} \label{eq:tensorattraction}, 
\end{eqnarray}
where $|^{3}S_{1}\rangle$ and $|^{3}D_{1;m}\rangle$ are wavefunctions of a proton-neutron pair in the $^{3}S_{1}$ and $^{3}D_{1;m}$ channels. $E_{0}$ and $E_{m}$ denote the energies of the unperturbed $^{3}S_{1}$ state and those of the virtually-excited $^{3}D_{1;m}$ states, while ${\cal Q}$ represents the Pauli blocking in the intermediate state, which is 0 when the state is occupied by other nucleons and 1 when it is not occupied.  In the case of deuteron, ${\cal Q}$ is always unity, but it is much smaller than unity below the Fermi surface in finite nuclei and in nuclear matter.  When the density is low, a proton (red) and a neutron (blue) in the left part of Fig.~\ref{fig:saturation} can be easily mixed with unoccupied states because the distance between the unperturbed state and the Fermi surface is small. However, the energy difference between the deeply bound particles and the Fermi surface increases as the density increases, resulting in an increase of the energy denominator in Eq.~(\ref{eq:tensorattraction}) and the second-order perturbation energy $\Delta E$ decreases. This is a major origin of the nuclear saturation at $\rho\sim 0.17$~fm$^{-3}$~\cite{Bethe71}.

\section{Spin effects in exotic nuclei -- change in spin-orbit coupling --}
As shown below, the spin-orbit coupling discussed in the previous section is subject to substantial changes in nuclei far from the $\beta$-stability line. 

\begin{figure}[htbp]
 \centering
 \resizebox{0.7\textwidth}{!}{\includegraphics{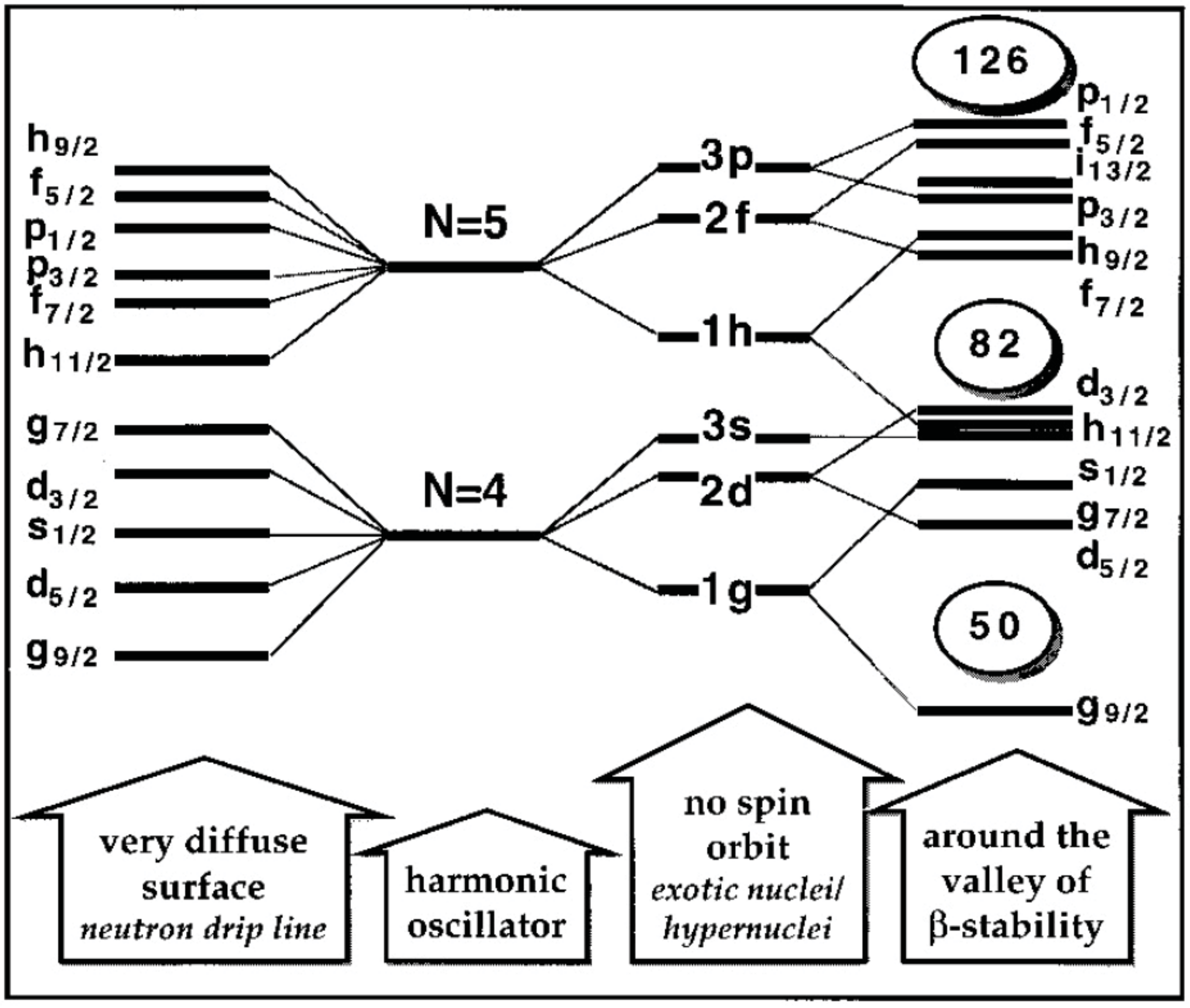}}
 \caption{(color online) Mean-field prediction of the change in the spin-orbit coupling. Reprinted from Ref.~\cite{Dobaczewski96}\copyright 1996 with permission from the American Physical Society. \label{fig:Dobaczewski}}
\end{figure}

One of the early works is a prediction by Dobaczewski and his collaborators, which indicates that the spin-orbit coupling may be significantly weakened in neutron-rich nuclei~\cite{Dobaczewski94} (Fig.~\ref{fig:Dobaczewski}). 
Several other theoretical works~\cite{Lalazissis98,Pudliner96} also predicted that changes in the spin-orbit splitting can take place in the region far from the stability, significantly affecting the nuclear structure.

\begin{figure}[htbp]
 \centering
 \resizebox{0.5\textwidth}{!}{\includegraphics{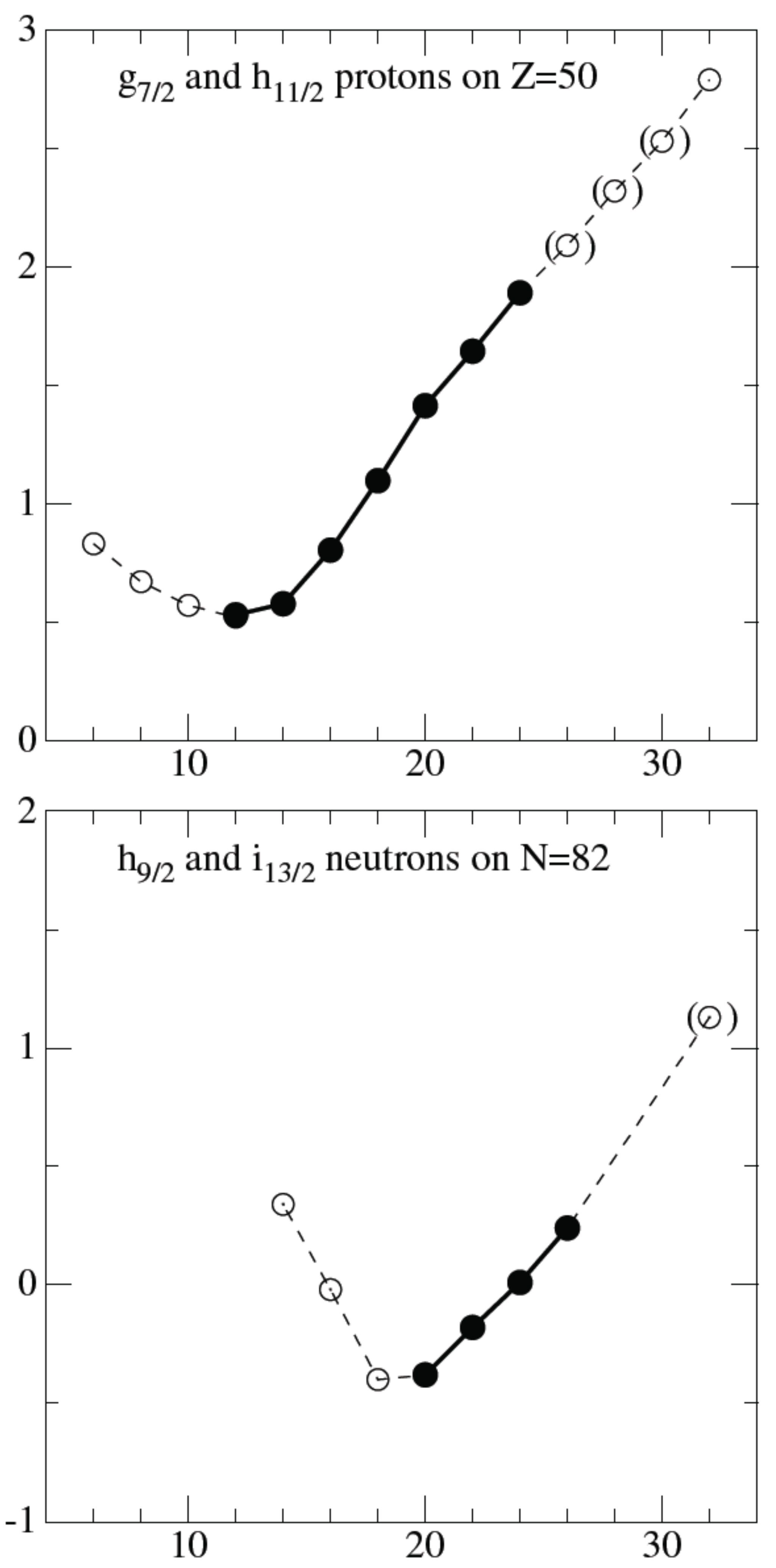}}
 \caption{Energy differences (in MeV) between the proton $h_{11/2}$ and $g_{7/2}$ orbits on the $Z=50$ shell (top) and the neutron $i_{13/2}$ and $h_{9/2}$ orbits on the  $N=82$ shell (bottom), as a function of the neutron excess of the core.
 Solid circles indicate information available from transfer reactions. Open circles represent assignments with methods which are less sensitive to the single-particle properties. Circles with parentheses are data with less certainty or by indirect assignments. 
 Reprinted from Ref.~\cite{Schiffer04} \copyright 2004 with permission from the American Physical Society.  \label{fig:Schiffer}}
\end{figure}

In 2004, Schiffer and his collaborators posed a question, ``Is the nuclear spin-orbit interaction changing with neutron excess?'' by reporting the experimental results on the energy difference between the proton $h_{11/2}$ and $g_{7/2}$ orbits~\cite{Schiffer04}. The interval between $h_{11/2}$ and $g_{7/2}$ increases as the spin-orbit splitting in the $h$- and/or $g$-orbits decreases and can be a good measure of the spin-orbit coupling.  The energy interval was determined by measuring the corresponding single particle energies with  the $(\alpha, t)$ reactions on Sn isotopes. The results shown in the upper panel of Fig.~\ref{fig:Schiffer} clearly indicate that the energy interval actually increases in the region of ${\rm ^{112-124}Sn}$. A similar behavior was also found in the neutron $i_{13/2}$ and $h_{9/2}$ orbits in the  $N=83$ isotones (lower panel of Fig.~\ref{fig:Schiffer}). In Ref.~\cite{Schiffer04}, it is claimed that ``{\it when beams of radioactive nuclei become available with sufficient intensity these trends can be explored further.}''

Otsuka and his collaborators proposed that the tensor force is essential to understand the shell evolution in the region far from the $\beta$ stability line~\cite{Otsuka05}, and specifically can explain the change in the proton $h_{11/2}$ and $g_{7/2}$ energies in Sn isotopes as a function of the neutron excess~\cite{Otsuka06}. It should be noted that they were the first group to thoroughly discuss the first-order tensor force effects. The effect causes the effective interactions between protons with $j=\ell\pm 1/2$ and neutrons with $j'=\ell'\mp 1/2$ to be more attractive than those between protons with $j=\ell\pm 1/2$ and neutrons with $j'=\ell'\pm 1/2$. This feature decreases, for example, the proton (neutron) spin-orbit coupling as neutrons (protons) are added to an orbit with $j=\ell+ 1/2$.

Thus, the interplay among the mean-field effects, the first and second-order tensor force effects, and possibly the three-nucleon force effects may significantly change the spin-orbit coupling in nuclei far from the stability line. The experiments with stable targets~\cite{Schiffer04,Noro-ECT2010} show quite interesting indications on the possible changes of the spin-orbit coupling.

\begin{figure}[htbp]
   \centering
  \resizebox{0.6\textwidth}{!}{\includegraphics{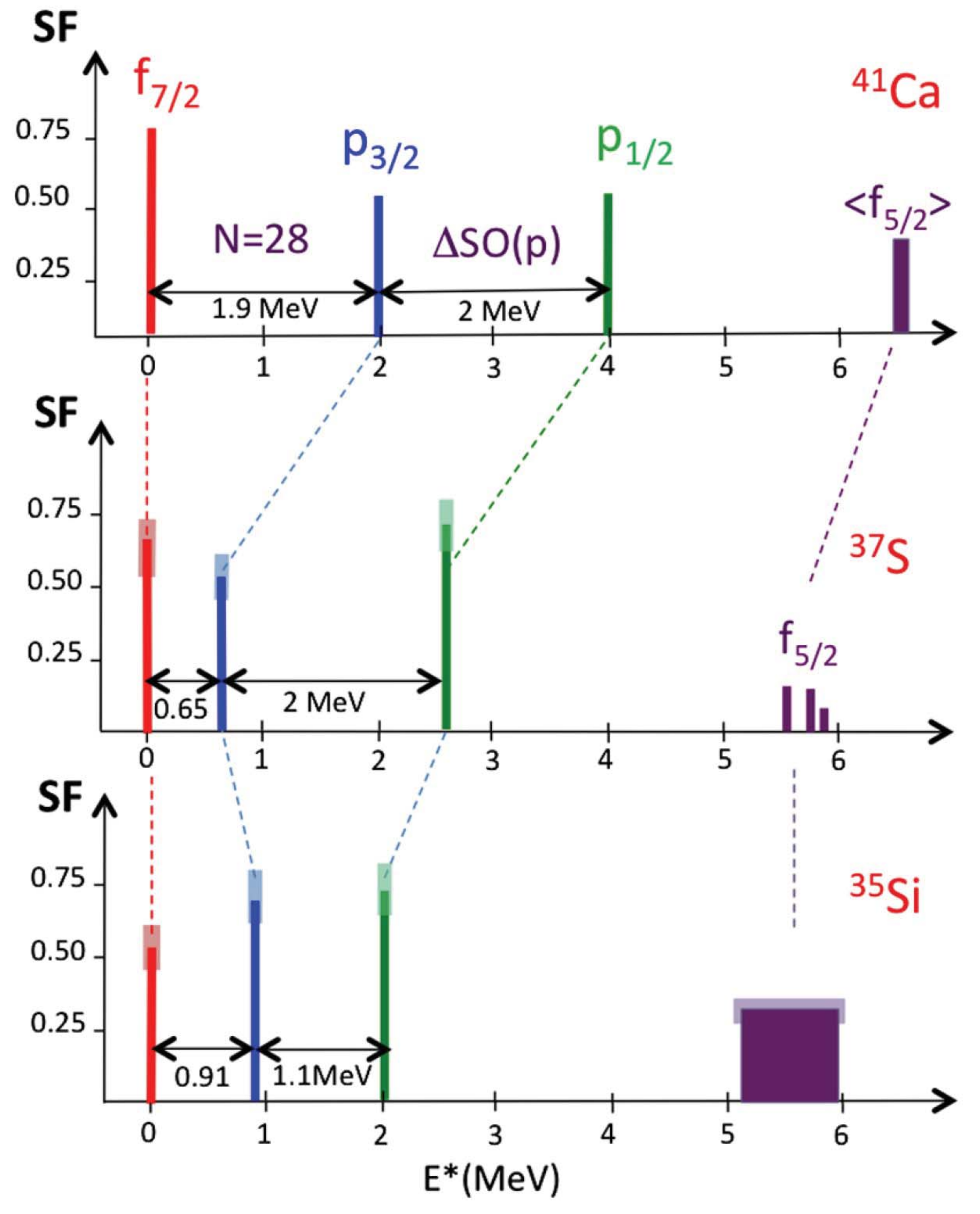}}
  \caption{(color online) Major single particle strength in ${\rm ^{41}Ca}$ (top), ${\rm ^{37}S}$ (middle), and ${\rm ^{35}Si}$(bottom).  Values of the spectroscopic factor in ${\rm ^{41}Ca}$  and the $5/2^{-}$ components in ${\rm ^{37}S}$ are taken from previous experiments, while all other spectroscopic factors are determined in Ref.~\cite{Burgunder14} with error bars due to statistics and fitting.
Reprinted from Ref.~\cite{Burgunder14} \copyright 2014 with permission from the American Physical Society. \label{fig:bubblenuclei}}
\end{figure}
One stimulating experimental result for unstable nuclei can be found in Ref.~\cite{Burgunder14}. In this experiment, neutron single particle energies in ${\rm ^{37}S}$ and ${\rm ^{35}Si}$ were determined with the $(d,p)$ neutron transfer reaction on the secondary ${\rm ^{36}S}$  and ${\rm ^{34}Si}$ beams produced at GANIL. The results together with those on ${\rm ^{41}Ca}$ are summarized in Fig.~\ref{fig:bubblenuclei}. The spin-orbit splitting between $1p_{3/2}$ and $1p_{1/2}$ is reduced in ${\rm ^{35}Si}$ by 25\% compared to that in ${\rm ^{37}S}$, while the change between $0f_{7/2}$ and $0f_{5/2}$ is insignificant.
One of the mechanisms to explain the changes is the occurrence of a dip in the density around the center of ${\rm ^{35}Si}$ due to the removal of two $1s_{1/2}$ protons from ${\rm ^{37}S}$, which creates a ``bubble'' at the center of  ${\rm ^{35}Si}$ and results in a reduction of the $1p_{3/2}$-$1p_{1/2}$ splitting without a significant effect on the $0f_{7/2}$-$0f_{5/2}$.

All of these experimental and theoretical studies have provided motivation to extend the experimental studies of the spin-orbit coupling with RI beams and polarized targets.

\section{Polarized targets for RI-Beam Experiments}
 
\subsection{Brief review of production of nuclear polarization}

Since the early 1950's, a variety of methods have been developed to polarize nuclei: atomic-beam methods~\cite{Haeberli67}, optical pumping methods~\cite{Happer72}, dynamical nuclear polarization  (DNP) methods~\cite{Goertz02,Crabb97}, brute force methods~\cite{Rose49,Honig76}, etc. 

The atomic-beam method was employed in the first polarized ion source constructed in Hamburg~\cite{Clausnitzer56}. Since then, polarized ion sources of this type have been constructed around the world and have provided polarized proton and deuteron beams for nuclear physics experiments. The same technique can be applied to a gas jet target system. At RHIC, a polarized gas-jet target of hydrogen is used in a polarimeter for a proton beam at relativistic energies~\cite{Zelenski05}. 

The optical pumping technique was applied to polarize a hydrogen gas jet and rare gases. In particular, spin-exchange~\cite{Bouchiat60} and metastability-exchange~\cite{Colegrove63} methods have been quite successful in polarizing rare gases such as ${\rm ^{3}He}$ and ${\rm ^{129}Xe}$, and have been used in nuclear and particle physics experiments as well as in medical applications~\cite{Albert94}.

Nuclei in solids can be polarized with DNP and brute force methods. While practical application of brute force methods are limited to cases of ferromagnetic atoms (Rose effect~\cite{Rose49}) and of the HD molecule~\cite{Honig76}, the DNP technique is more versatile: since its first application to nuclear and high-energy physics experiments~\cite{Abragam62,Chamberlain63}, many polarized proton/deuteron targets have been constructed based on the DNP technique~\cite{Goertz02,Crabb97}. In the traditional DNP method, electrons thermally polarized at a temperature less than 1~Kelvin and in a high magnetic field of several Tesla are used to polarize nuclei. Recently, a new technique, the so-called {\it triplet-DNP}, has been introduced.

\subsection{Requirements for RI-beam experiments \label{sec:reqpoltargets}}
\begin{figure}[htbp]
 \centering
 \resizebox{0.8\textwidth}{!}{\includegraphics{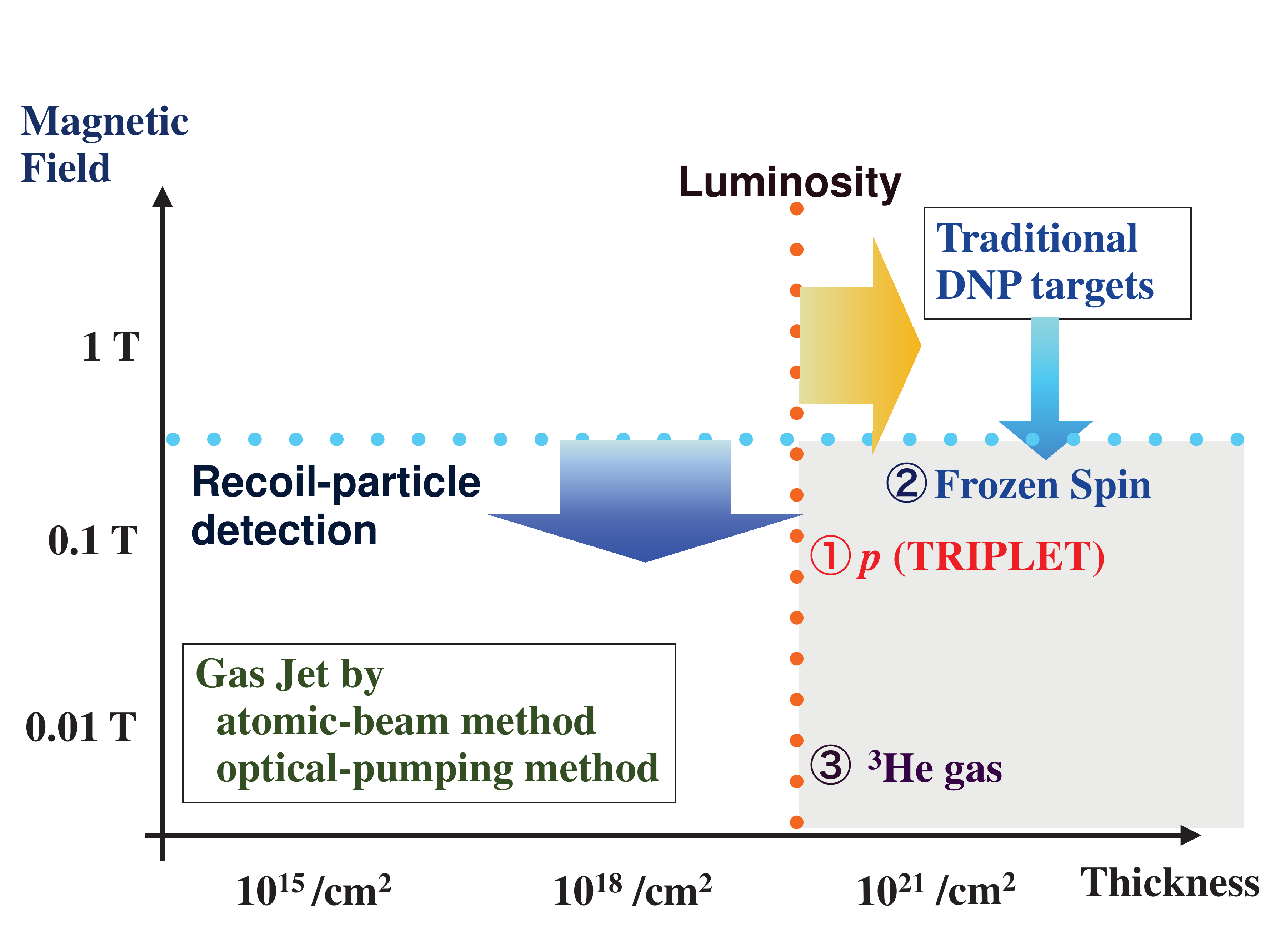}}
 \caption{(color online) Polarized targets in the market. Requirements for RI beam experiments are shown with dotted lines. Triplet-DNP, frozen-spin DNP, optically-pumped ${\rm ^{3}He}$ targets suffice the requirements on the magnetic field strength and on the target thickness in order to reach a reasonable luminosity of $> 10^{23}$~s$^{-1}$cm$^{-2}$ for typical beam intensities.  \label{fig:PolTargets}}
\end{figure}

RI beam experiments impose requirements different from those in other applications~\cite{Uesaka04,ObertelliUesaka11}. Figure~\ref{fig:PolTargets} summarizes polarized targets on the market and requirements for RI-beam experiments (dotted lines), discussed below.

RI beams produced by nuclear reactions usually have a low intensity. To achieve significant luminosity with the low-intensity secondary beams, the target should be sufficiently thick. In the following, for quantitative discussion, it is assumed that a beam intensity is 10$^{3}$~s$^{-1}$. When one requires that the luminosity is larger than 10$^{23}$~s$^{-1}$cm$^{-2}$, which corresponds to a counting rate of $\sim$100~counts/day for a cross section of 10~mb, a criterion for the target thickness is $> 10^{20}$~cm$^{-2}$.  This criterion is indicated with a vertical dotted line in Fig.~\ref{fig:PolTargets}. A gas jet target with a thickness of 
  $\sim 10^{11}$~cm$^{-2}$ does not meet the requirements, except in the case of storage ring experiments. On the other hand, solid targets based on DNP techniques and ${\rm ^{3}He}$ targets based on the optical pumping techniques do suffice. The following discussion is devoted only to DNP targets. A case of the ${\rm ^{3}He}$ target is discussed in Section~\ref{sec:active3He}.

Any spin-polarized solid target is a compound of hydrogen and other elements. Thus, it is necessary to identify reactions with hydrogen from those with other contaminant atoms in the target. In many cases, detection of the recoil protons facilitates the event identification. The recoil protons scattered at forward angles in the center-of-mass system have an energy as low as 5--10 MeV. Thus the external magnetic field, which is indispensable in polarized targets, should be as low as 0.3~T to enable low-energy proton detection. This criterion is shown in Fig.~\ref{fig:PolTargets} with a horizontal dotted line. 

Low magnetic field operations are needed even when the recoiled particles have a sufficiently large energy, but the requirement for the angular resolution is high. This is the case in $(p,pN)$ experiments in inverse kinematics condition, where an angular resolution of a few milliradian is required. 

Because traditional dynamic-nuclear-polarization (DNP) targets~\cite{Goertz02,Crabb97}
 use thermal polarization of electrons to polarize protons, a magnetic field higher than a few Tesla and a sub-Kelvin temperature are necessary. For this reason, it is difficult to apply a DNP target to RI-beam experiments as it is. 

There are two possibilities to enable low-magnetic field operations: a spin-frozen operation of traditional DNP targets and use of a triplet-DNP target where protons are polarized by transferring an electron polarization in a photo-excited triplet state of an aromatic molecule. As of today, the latter technique has been used in a target for experiments with a RI beam. The next subsection overviews the triplet-DNP technique.

\subsection{Triplet-DNP target}
\subsubsection{Overview of triplet-DNP target}
To produce proton polarization under a low magnetic field, non-thermal polarization of electrons must be used. A large population difference among Zeeman sub-levels, hereinafter referred to as ``electron polarization'',  in the photo-excited triplet state of an aromatic molecule~\cite{Kesteren85,Henstra90} is a promising candidate. Proton polarization can be produced by transferring the electron polarization by means of a cross polarization technique~\cite{Henstra90}. The value of the electron polarization depends on neither its temperature nor the strength of the external magnetic field, which makes it possible to polarize protons up to several tens of percent~\cite{Iinuma00,Wakui05,Takeda02} at a temperature higher than 77 K and in a magnetic field lower than 0.3~T. 

The first polarized proton target for RI-beam experiments was constructed based on the triplet-DNP technique~\cite{Uesaka04}. The target material was a single crystal of naphthalene (C$_{10}$H$_{8}$) doped with a small amount (0.005 mol\%) of pentacene (C$_{22}$H$_{14}$). The target was formed through the Bridgemann crystallization after zone melting refinement. For use in scattering experiments, the crystal is shaped into a 1-mm-thick disk. To cover the spot size of RI-beams, the diameter of the cross section is as large as 14~mm.
The target holder is composed of hydrogen-free polychlorotrifluoroethylene to avoid background production.

Electron polarization of pentacene produced by irradiating with a 515-nm light from Ar-ion lasers~\cite{Wakui05} is
transferred to protons with the cross polarization method~\cite{Henstra90}. In this method, a microwave at the electron spin resonance frequency is irradiated to the target while an external magnetic field is varied simultaneously to cover the inhomogeneously broadened line-width. To maintain the high penetrability of the recoiling protons, a thin copper film loop-gap resonator (LGR)~\cite{Ghim96} is used as the microwave resonator. The copper film LGR is a thin Teflon cylinder with copper platings on both sides. The resonant frequency is determined by plating geometry, the Teflon thickness, and the resonator length. The thicknesses of the Teflon sheet and copper plating are 25~$\mu$m and 4.4~$\mu$m, respectively. The actual resonant frequency is 2--3 GHz depending on the design.

The first experiments with the target system were to measure the vector analyzing power for the $p$-${\rm ^{6,8}He}$ scattering at the RIPS facility, RIKEN~\cite{Uesaka10,Sakaguchi11,Sakaguchi13}. In these experiments, the target was operated in a low magnetic field of about 0.1~T and at temperature of 100~K. The proton polarization was in the range of 10--20\%. The experimental details are shown in Section~\ref{sec:lscoupingreaction}.

\subsubsection{Progress in Triplet-DNP target development}

Currently, proton polarization is limited by a lack of intense laser light to excite pentacene. 
 To achieve higher proton polarization in a material with a reasonably large volume, a high-power laser that satisfies the requirements summarized in Table~\ref{tbl:laser} is needed. Although different lasers, an Ar-ion laser~\cite{Wakui05}, a dye laser~\cite{Iinuma95}, and a disk laser~\cite{Eichhorn14} have been tested, none satisfies all the requirements.
\begin{table}[htbp]
 \centering
 \caption{Specifications required for a laser to be used in pentacene excitation\label{tbl:laser}}
 \begin{tabular}{lr}
 \hline
 Wavelength [nm]        & 590, 545, or 510 \\
 Pulse width [$\mu$s]   & $\sim$1 \\
 Repetition rate [Hz]   & $\sim$ 1000 \\
 \hline
 \end{tabular}
\end{table}

\begin{figure}[htbp]
  \centering
  \resizebox{0.55\textwidth}{!}{\includegraphics{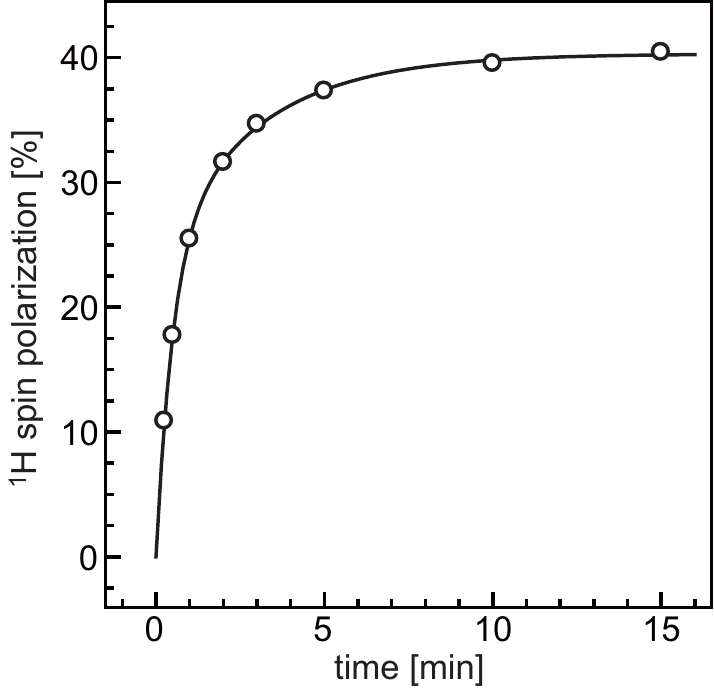}}
  \caption{Proton polarization at 0.6~T and at room temperature. A new 589-nm laser with a pulse width of 126~ns and a repetition rate of 1~kHz is used. \label{fig:pol}}
\end{figure}

The RIKEN group has started to construct a new laser system which satisfies the requirements and fits best to the pentacene excitation. This laser system consists of LD-driven YAG lasers with output wavelength of 1064~nm and 1319~nm and a sum frequency generator crystal to convert infrared light into 589-nm light which has a sum energy of the 1064- and 1319-nm photons. Test experiments by a prototype laser are found to be quite promising~\cite{Uesaka-EURISOL}: a proton polarization of $\sim$41~\% has been achieved at 0.6~T and room temperature in a sample of $p$-terphenyl-$d_{4}$ doped with pentacene-$d_{14}$ for a 126-ns pulse width of and a 1-kHz repetition rate (Fig.~\ref{fig:pol}).

The new direction to produce a high proton polarization in solid materials at {\it room temperature} will definitely expand the applicability of proton polarization. One promising direction is to apply a polarized target to low-energy RI-beam experiments. As discussed in Sections~\ref{sec:resonantscattering} and \ref{sec:transfer}, polarization observables in low-energy reactions such as transfer reactions and proton resonant scatterings are good signatures of the spin-dependent nuclear structure. Such sensitivities are hardly available otherwise. In the presently used setup, the target material is cooled to $\sim$100~K by keeping it in cold nitrogen gas atmosphere. The room-temperature operation removes nitrogen gas ($\sim$40~mg/cm$^2$) and films to separate the beam-line vacuum which are serious drawbacks in low-energy experiments~\cite{Sakaguchi13}. 

To achieve a high proton polarization at room temperature, the nuclear spin relaxation rate must be suppressed and/or the photo-excitation efficiency increased. It should be noted that, employing $p$-terphenyl ($\rm C_{18}H_{14}$), instead of naphthalene is beneficial at room-temperature because pentacene in a $p$-terphenyl crystal can be doped with a one-order of magnitude higher concentration ($\sim$0.1~mol\%) than in naphthalene, which results in a higher efficiency in polarizing protons as long as sufficiently intense laser is available.

Recently, Tateishi and his collaborators reported that they successfully suppressed the nuclear spin relaxation at room temperature by (partially) deuterating  $p$-terphenyl and pentacene~\cite{Tateishi14}. They replaced 4 of the 14 hydrogen atoms in a $p$-terphenyl ($p$-terphenyl-$d_{4}$) and all of the hydrogen atoms in pentacene (pentacene-$d_{14}$) by deuterium. The spin-relaxation rate is reduced by a factor of $\sim$4. For the partially deuterated sample, they achieved a proton polarization of 34\%.

\section{Spin-orbit coupling in nuclear reaction \label{sec:lscoupingreaction}}
The triplet-DNP target was used to experimentally measure the vector analyzing power for the $p$-${\rm ^{6,8}He}$ scattering at RIKEN~\cite{Uesaka10,Sakaguchi11,Sakaguchi13}.  As discussed in Section~\ref{sec:spinorbithistory}, the vector analyzing power in elastic scattering of non-zero spin particle is a direct manifestation of the spin-orbit coupling in nuclei.

The radial shape of the spin-orbit potential can be approximated with a form  $V_{\rm \ell s} = \frac{1}{r}\frac{d\rho(r)}{dr}$. Here $\rho(r)$ is the nuclear density at a radius $r$.  Since the nuclear density is almost constant in the interior of a nucleus, the spin-orbit potential is localized in the vicinity of the nuclear surface. From the surface nature of the spin-orbit potential, it is readily expected that different distributions
of protons and neutrons in neutron-rich nuclei may modify the shape. In particular, 
when neutrons have an extended distribution as is typical in
skin or halo nuclei, the spin-orbit potential should have a corresponding extended shape. This possibility can be experimentally investigated with a polarized proton target and beams of neutron-rich nuclei.

As the first physics case, neutron-rich ${\rm ^{6,8}He}$ nuclei were employed. It was expected that the distinctive neutron distributions in  ${\rm ^{6,8}He}$ would modify the magnitude and radial shape of the spin-orbit potential.

\begin{figure}[htbp]
 \centering
  \resizebox{0.7\textwidth}{!}{\includegraphics{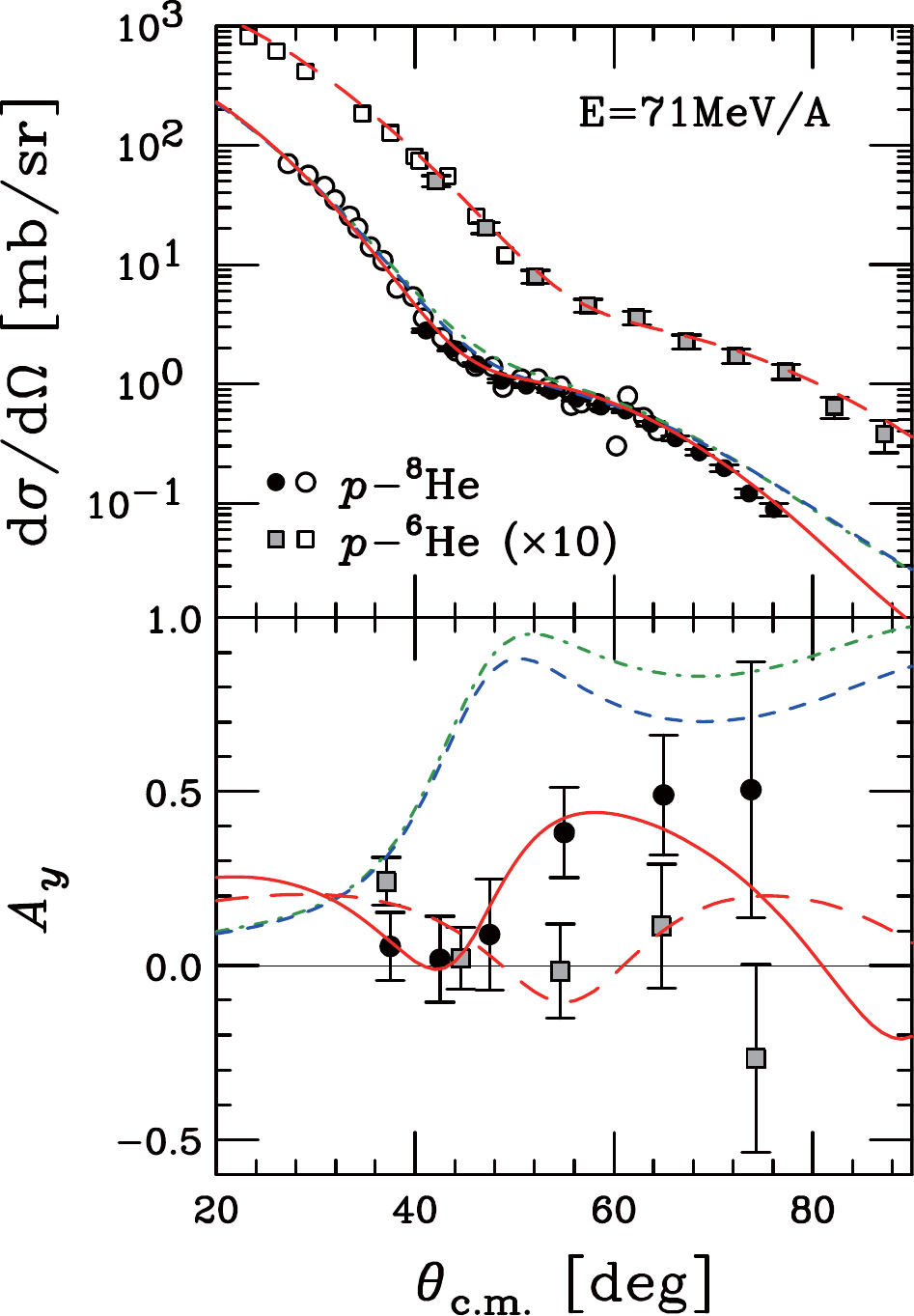}}
  \caption{(color online) Cross section (upper) and vector analyzing power (lower) for the $p-{\rm ^{6,8}He}$ elastic scattering at 71~MeV/u. Reprinted from Ref.~\cite{Sakaguchi13}\copyright 2013 with permission from the American Physical Society. \label{fig:p68He}}
\end{figure}

${\rm ^{6,8}He}$ beams at an energy of 71 MeV/u and an intensity of $\sim$2$\times 10^5$ pps produced at the RIPS facility~\cite{Kubo92} of RIKEN have been used to bombard the polarized target. Low magnetic field operations allow the low-energy recoiling protons to be detected and the reaction events to be identified.
Data of the vector analyzing power, together with the cross section, were measured for the elastic scattering of polarized proton and ${\rm ^{6,8}He}$  in the angular range of 39--78$^{\circ}$ in the center of mass system~\cite{Uesaka10,Sakaguchi11,Sakaguchi13}. 

Figure~\ref{fig:p68He} shows the data of the cross section and the vector analyzing power where filled circles, shaded squares, and open circles/squares are from Refs.~\cite{Sakaguchi13}, \cite{Uesaka10,Sakaguchi11}, and  \cite{Korsheninnikov93,Korsheninnikov97}, respectively.  The solid and long-dashed curves in Fig.~\ref{fig:p68He} are the best-fit results for $p-{\rm ^{8}He}$ and $p-{\rm ^{6}He}$, while the short-dashed and dot-dashed lines are the results with the same central terms for ${\rm ^{6}He}$ but with the spin-orbit terms of global optical potentials,  KD03~\cite{KoningDelaroche03} and CH89~\cite{Varner91}. The spin-orbit terms of the global potentials are clearly incompatible with the analyzing power data although the effect is negligibly small in the cross section.

\begin{figure}[htbp]
 \centering
  \resizebox{0.7\textwidth}{!}{\includegraphics{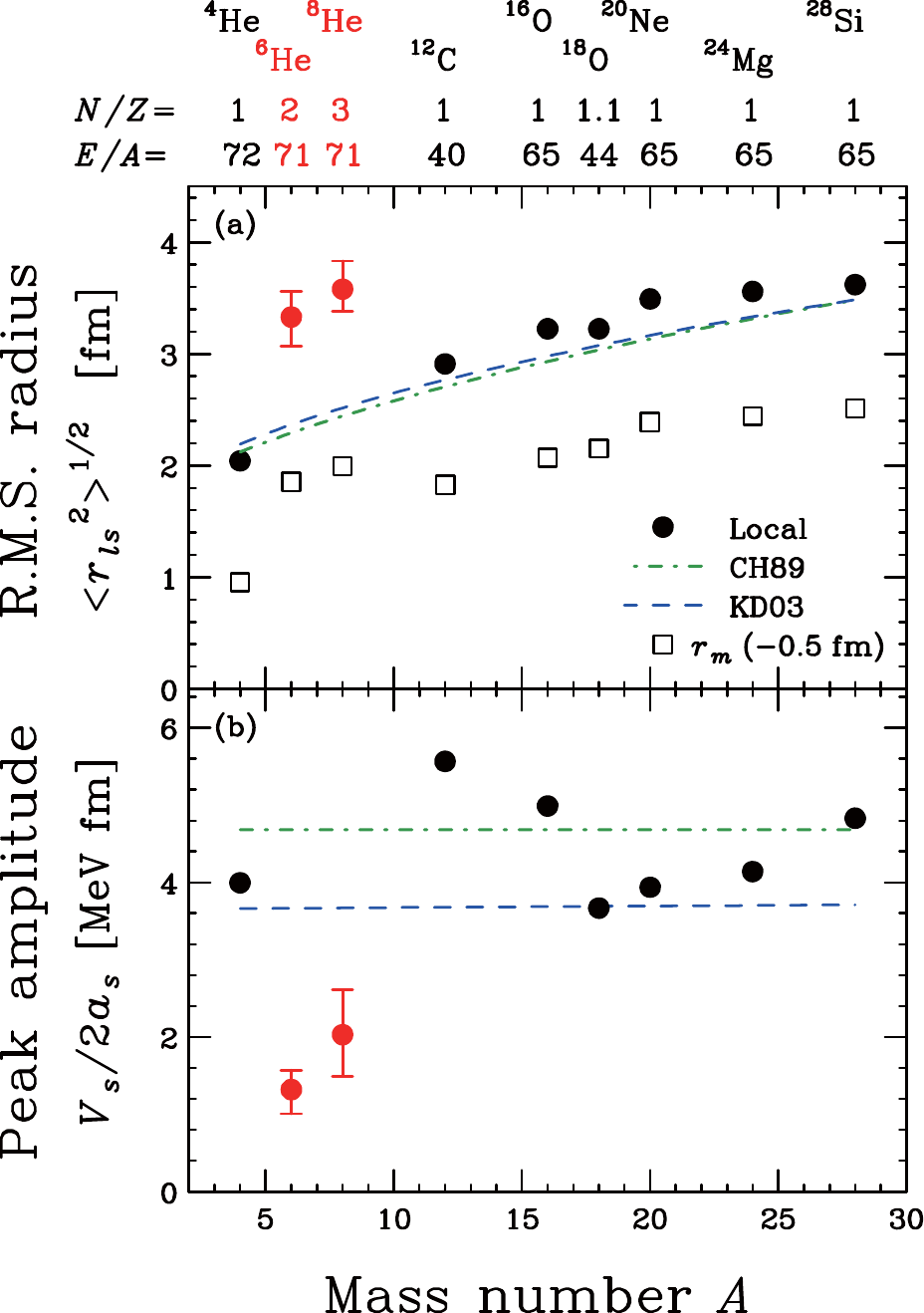}}
  \caption{(color online) Root mean square radius $\langle r_{\ell s}^2\rangle^{1/2}$ and the peak amplitude $V_{s}/2a_{s}$ of the spin-orbit potential. Reprinted from Ref.~\cite{Sakaguchi13}\copyright 2013 with permission from the American Physical Society.  \label{fig:p68HeElastic}}
\end{figure}

To clarify the difference between the optical potentials that reproduce the analyzing power data and the global optical potential, Fig.~\ref{fig:p68HeElastic} plots the root mean square radius $\langle r_{\ell s}^2\rangle^{1/2}$ and the peak amplitude $V_{s}/2a_{s}$ of the spin-orbit potential~\cite{Sakaguchi13}, where $V_{s}$ and $a_{s}$ are depth and diffuseness parameters of the spin-orbit potential, respectively. It should be noted that $V_{s}$ is defined in the unit of MeV$\cdot$fm$^2$ (see Ref.~\cite{Sakaguchi13} for detail). The root mean square radii of the ${\rm ^{6,8}He}$ spin-orbit potentials are significantly larger than those of stable nuclei and the global potentials. On the other hand, the peak amplitudes in ${\rm ^{6,8}He}$ are as small as 1.3--2~MeV$\cdot$fm, while those in stable nuclei are in the range of 3.5--5.5~MeV$\cdot$fm. Thus, it is concluded that the spin-orbit potentials between a proton and neutron-rich ${\rm ^{6,8}He}$ nuclei are shallower and more diffuse than the global
systematics of nuclei along the stability line. This is considered to be a consequence of the diffuse density distribution in the neutron-rich isotopes.

\section{Experimental approaches to spin-orbit splitting in exotic nuclei}
An energy splitting between a spin doublet, which is known as {\it spin-orbit splitting}, is 
a good measure of the spin-orbit coupling in nuclei. Experimental determination of the spin-orbit splitting demands reactions with the following two features: selectivity toward single particle or hole states and sensitivity to the total angular momentum, $j$. In this section, reactions with these features, including the $(p,pN)$ quasi-free scattering, the proton resonant scattering, and transfer reactions are discussed, by emphasizing the spin observables. 

\subsection{$(p,pN)$ Quasi-free scattering}

At energies of several hundreds MeV/u, quasi-free knockout $(p,pN)$ reactions can be a good spectroscopic tool to study single hole states~\cite{JacobMaris66,JacobMaris73}.
The reaction is dominated by a single step nucleon-nucleon scattering in nuclear medium
and it selectively populates single hole states without seriously disturbing the residual nucleus. Consequently, reaction analyses based on distorted-wave impulse approximation (DWIA) can be reliably used at these energies.  The missing momentum dependence of the cross section corresponds directly to the momentum distribution of the knocked-out nucleon, allowing the orbital angular momentum $\ell$ and the spectroscopic strength of the orbit to be determined. 
The recent experimental work at the Research Center of Nuclear Physics (RCNP), Osaka University, demonstrated that the spectroscopic factors determined by DWIA analyses of $(p,2p)$ reaction at $E_p \sim$200~MeV are consistent with those by $(e,e'p)$ reactions~\cite{Noro-ECT2010}. Although uncertainty in the optical model parameters used in the reaction analysis of unstable nuclei might prevent the absolute values of the spectroscopic factors from being determined, the relative values should be free from serious uncertainties. 

\begin{figure}[htbp]
   \centering
  \resizebox{0.6\textwidth}{!}{\includegraphics{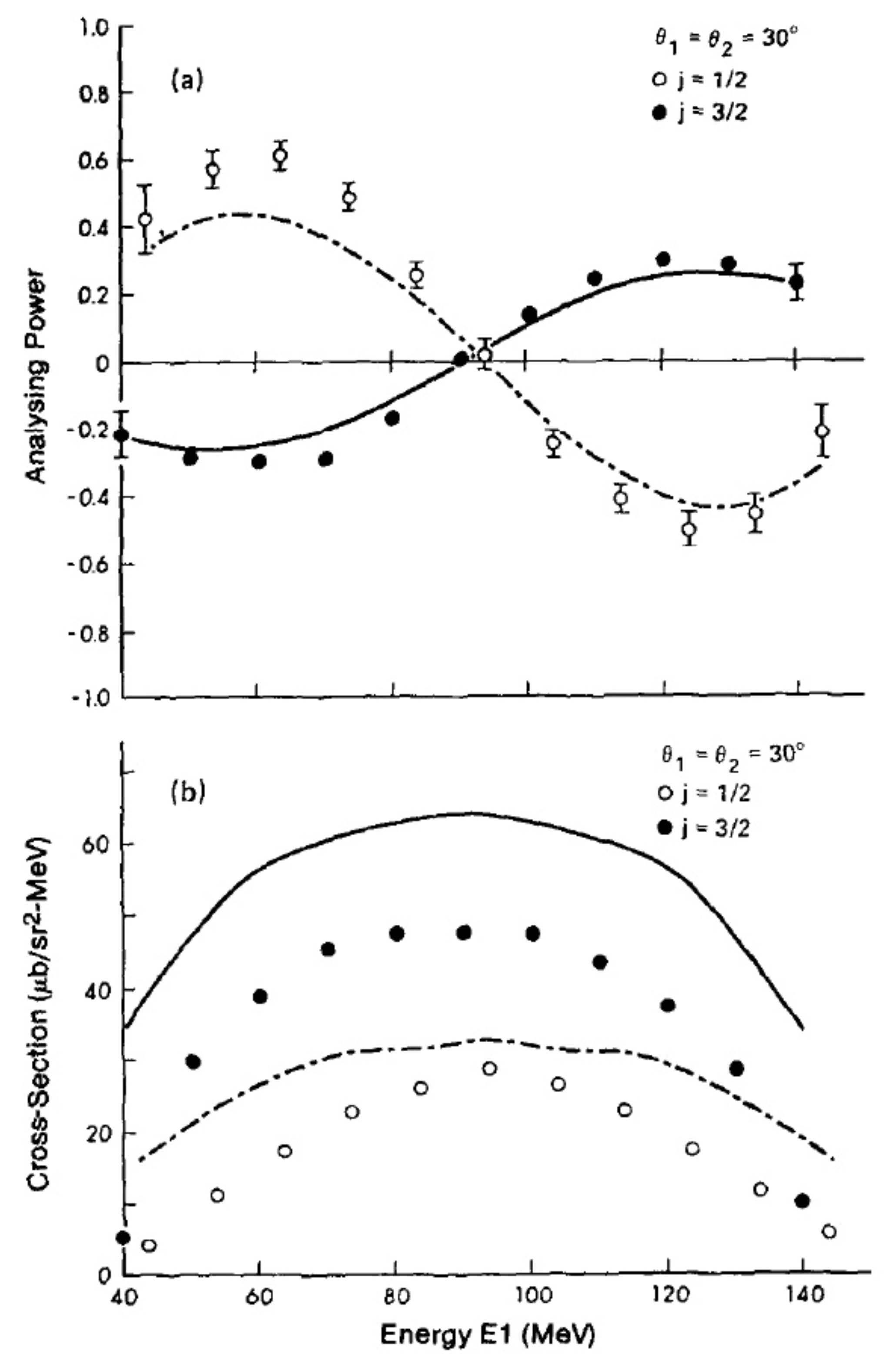}}
  \caption{Analyzing power and cross section for the ${\rm ^{16}O}(p,2p)$ reaction at 200~MeV. Reprinted from Ref.~\cite{Kitching80}\copyright 1980 with permission from Elsevier. \label{fig:Kitching}}
\end{figure}

To equip the spin ($j$) determination capability to the $(p,pN)$ reaction, we need polarized protons.
The principle of the $j$-determination by the  $(p{\rm (pol)},pN)$ reaction proposed theoretically by Jacob and Maris~\cite{Jacob73,Jacob76} is particularly effective in an incident energy region of 200--400~MeV where the spin correlation parameter $C_{nn}$ in the nucleon-nucleon scattering is large. 

Kitching and his collaborator performed the ${\rm ^{16}O}(p,2p)$ experiment at TRIUMF  by using a 200-MeV polarized proton beam~\cite{Kitching80,Kitching76}. Two protons in the final state were detected at 30$^{\circ}$ in the laboratory system. Figure~\ref{fig:Kitching} shows data of the vector analyzing power (a) and the cross section (b). Filled and open circles represent data for knockout of protons in the $p_{1/2}$ and $p_{3/2}$ states, respectively, as a function of the kinetic energy of one proton $E_{1}$. The $E_{1}$ dependence of the cross sections shown in Fig.~\ref{fig:Kitching}~(b)  is almost identical for the $p_{1/2}$ and $p_{3/2}$ knockouts, clearly demonstrating that a cross section measurement alone does not allow $j$ to be determined. In contrast, because the sign of the vector analyzing power data depends on the $j$, it can provide a clear signature of $j$.

The $(p,2p)$ knockout reactions have been recently extended to RI beam experiments with unpolarized~\cite{Panin16} and polarized~\cite{Kawase15,Tang15} targets.

\subsection{Proton resonant scattering\label{sec:resonantscattering}}
Proton resonant scattering is a known tool to investigate single particle states of protons at low incident energies. It can probe resonant states of a proton and a heavy ion in the relative energy region of several MeV. This technique is particularly efficient when applied to experiments with RI-beams: a thick target method allows a relatively thick proton target of a few mg/cm$^2$ to be used for low-energy RI-beams~\cite{Goldberg93,Charity16}. 

\begin{figure}[htbp]
   \centering
  \resizebox{0.7\textwidth}{!}{\includegraphics{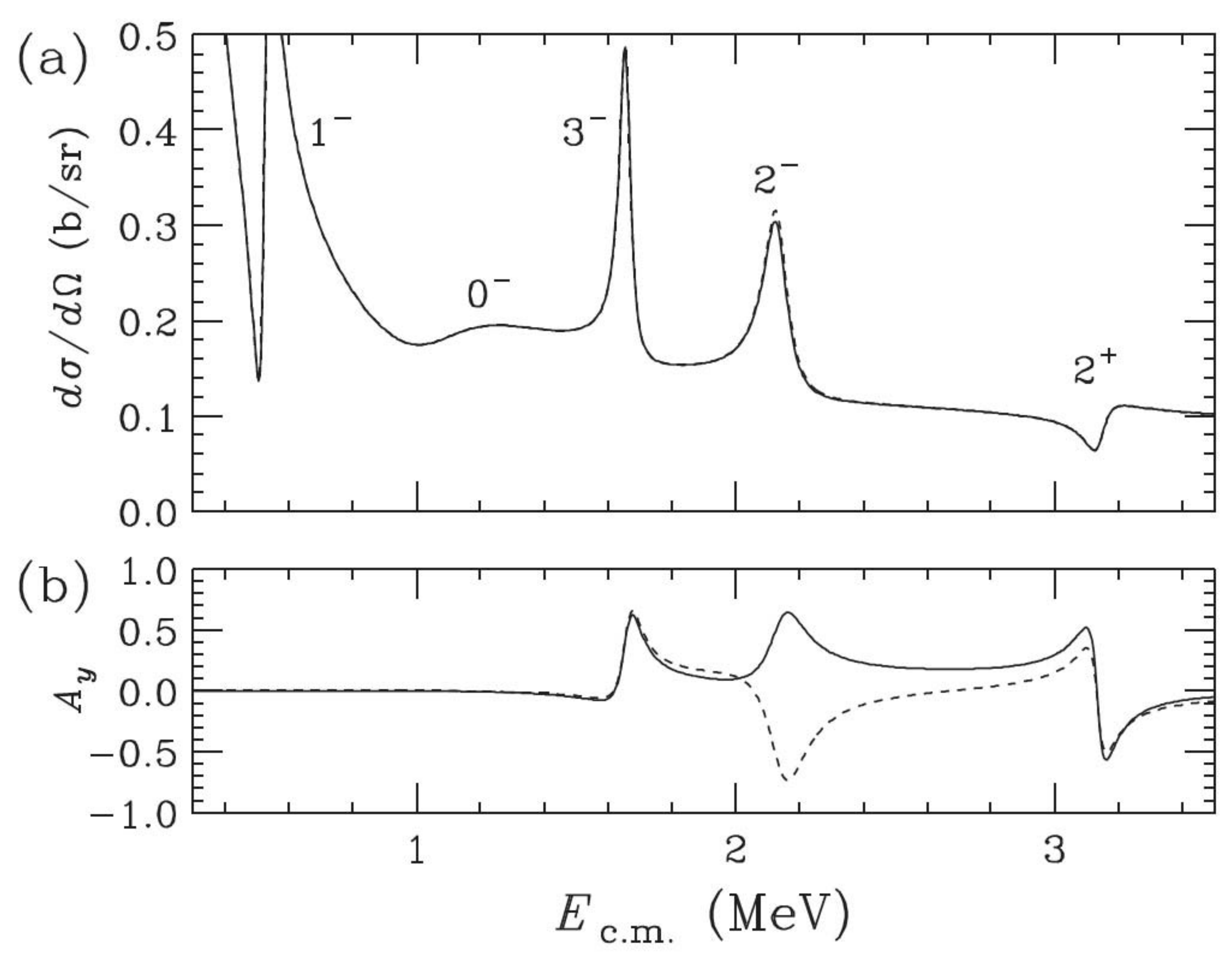}}
  \caption{R-matrix simulation of the analyzing power for the $p-{\rm ^{13}N}$ resonant elastic scattering. Reprinted from Ref.~\cite{Teranishi13} \copyright 2013, with permission from the American Institute of Physics. \label{fig:TeranishiRmatrix}}
\end{figure}

Measurements of the vector analyzing power will offer new opportunities~\cite{Teranishi13}. Teranishi demonstrated the new capabilities by means of R-matrix calculations~\cite{Teranishi13}. Figures~\ref{fig:TeranishiRmatrix} (a) and (b) show the cross section
and $A_{y}$ for the ${\rm ^{13}N}+p$ resonant scattering. The calculation is made for the center-of-mass scattering angle of 135$^{\circ}$. Since the spin-parity of ${\rm ^{13}N}$ is 1/2$^{-}$, low-lying resonances in ${\rm ^{14}O}$ are 0$^{-}$ and 1$^{-}$ (s-wave) and 3$^{-}$ and 2$^{-}$ (p-wave).  The analyzing power, $A_{y}$, is almost zero in the region where the s-wave resonances dominate, but takes finite values at larger excitation energies. There are two possible proton single particle orbits for the $2^{-}$ resonances: $d_{5/2}$ and $d_{3/2}$, which are plotted with solid ($d_{5/2}$) and dashed ($d_{3/2}$) lines. It is striking that $A_{y}$ takes the opposite sign depending on the proton configuration, while there is no obvious difference in the cross section. Thus $A_{y}$ can be used to clarify the proton single particle configuration. Reference~\cite{Teranishi13} pointed out that $A_{y}$ can be also powerful in decomposing resonances, especially when wide resonances overlap.

\subsection{Transfer reactions\label{sec:transfer}}
Transfer reactions are the most established experimental approaches to single particle states in nuclei.  In particular, this is one of the best playground for polarized beams. A good example can be found in the $(d,p)$ reaction studies on single neutron states~\cite{YuleHaeberli67,YuleHaeberli68}. It was already known that the angular distribution of the cross section can be used to determine the orbital angular momentum $\ell$ of the
captured neutron. This means, on the other hand, states with the same $\ell$ but with different values of total angular momentum $j=\ell \pm 1/2$ cannot be differentiated only with the cross section.

\begin{figure}[htbp]
   \centering
  \resizebox{0.6\textwidth}{!}{\includegraphics{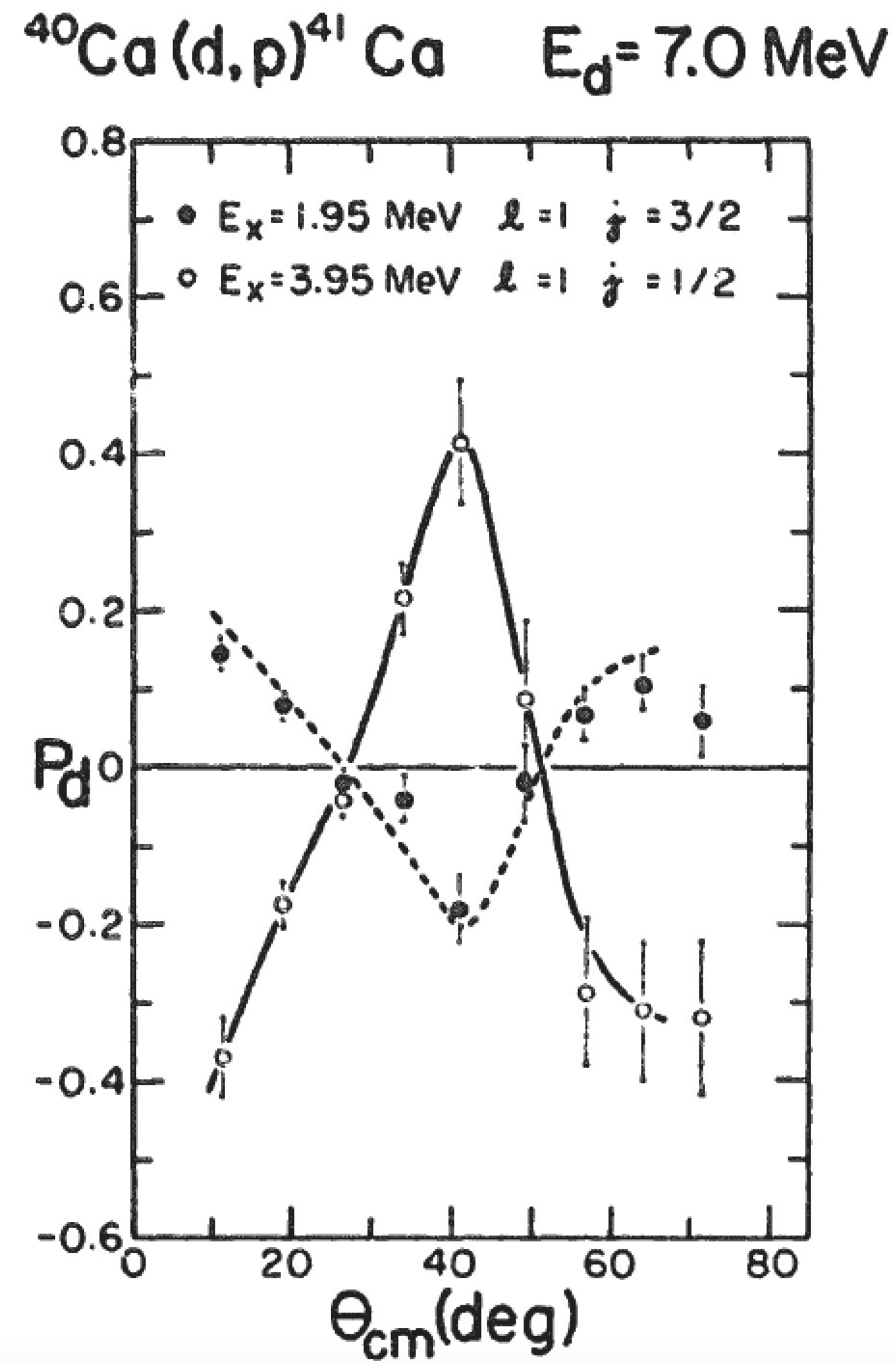}}
  \caption{Analyzing power for the ${\rm ^{40}Ca}(d,p)$ reaction at $E_{d}=7$~MeV. Reprinted from Ref.~\cite{YuleHaeberli68}\copyright 1968 with permission from Elsevier. 
  \label{fig:YuleHaeberli}}
\end{figure}

Figure~\ref{fig:YuleHaeberli} shows the vector analyzing power measured for the ${\rm ^{40}Ca(d,p)^{41}Ca}$ reaction at an incident deuteron energy of 7.0~MeV~\cite{YuleHaeberli68}. In the experiment, single neutron orbits on top of the closed ${\rm ^{40}Ca}$ core were investigated.  The states at 1.95~MeV and 3.95~MeV in ${\rm ^{41}Ca}$ have neutrons in $1p_{3/2}$ and $1p_{1/2}$ orbits and the angular distributions of the cross section for the states have a similar shape. In contrast, the sign of the vector analyzing power shown in Fig.~\ref{fig:YuleHaeberli} depends on $j$. Additionally, the ratio of the magnitude is close to the value of ${-\ell}/(\ell+1)=-0.5$, which is inferred by a distorted wave Born approximation.

\section{Future polarized targets}

The triplet-DNP is not the unique technique to prepare polarized targets for RI-beam experiments. Different capabilities of various methods can bring further breakthroughs. The first promising method is the spin-frozen operation of traditional DNP targets. This possibility has been pursued by a group at the Oak Ridge National Laboratory (ORNL) and the Paul Scherrer Institute (PSI)~\cite{UrregoBlanco05}. Another example, which is extremely ambitious and fascinating, is an active target with a ${\rm ^{3}He}$ gas polarized by means of the spin-exchange or metastability exchange method. The use of a gas-jet proton/deuteron target polarized by the atomic-beam or the optical-pumping methods is another promising solution when used in storage-ring experiments. These methods are well established and recent devices can provide an atomic beam of polarized hydrogen/deuterium with a flux of 10$^{16-17}$ atoms/sec. 

The possibilities of the spin-frozen operation of DNP targets and an active target with a ${\rm ^{3}He}$ gas are discussed below. 

\subsection{Spin-frozen operation of DNP targets}
  State-of-the-art DNP techniques can be, in principle, applied to RI-beam experiments by introducing "spin-frozen" operation. A collaboration of the ORNL and the PSI has investigated this possibility~\cite{UrregoBlanco05,UrregoBlanco07}, where the target material is a polystyrene plastic (${\rm C_{8}H_{8}}$) with a free nitroxyl radical TEMPO~\cite{Bunyatova04}. Due to the flexibility of plastic, the target can be formed into any shape to fit the experimental requirements. The availability of a target with a thickness of less than 1~mg/cm$^2$ is a big advantage in applications to experiments with low-energy RI beams. The target film is cooled through a copper frame which is in thermal contact with the mixing chamber of a dilution refrigerator.  A cryogenic system composed of the ${\rm ^3He}$-${\rm ^4He}$ dilution refrigerator and of a ${\rm ^4He}$ cryostat is used to cool the copper frame and superconducting coils. It is also used to prepare superfluid liquid helium that is in direct thermal contact with the target film.

In a test experiment, a high magnetic field of 2.5~T was added on the target during the polarization mode, achieving an electron polarization of about 100\% at 0.2~K. Irradiation of 70-GHz microwaves with a modulation of 1~kHz induced polarization transfer from electrons to protons. A proton polarization of 30\% was achieved~\cite{UrregoBlanco07} when no beam was injected to the target.
\begin{figure}[htbp]
 \centering
  \resizebox{0.7\textwidth}{!}{
  \includegraphics{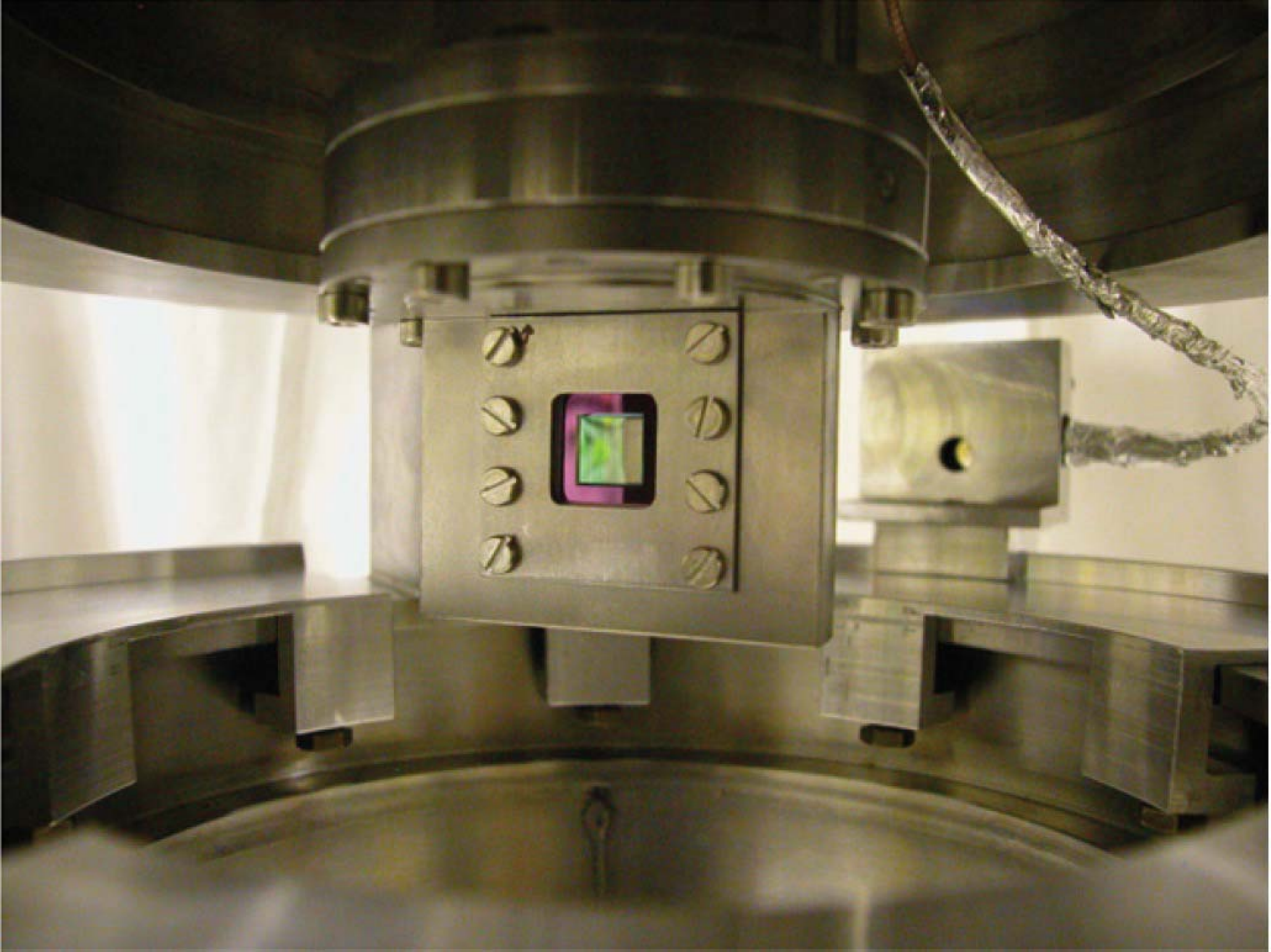}}
  \caption{(color online) Photo of ORNL-PSI polarized target. Reprinted from Ref.~\cite{UrregoBlanco07} \copyright 2007, with permission from Elsevier.  \label{fig:ORNLtarget}}
\end{figure}
In the frozen spin mode, the microwave irradiation was stopped and the magnetic field strength was reduced. Under this condition, the proton polarization survives on a time-scale defined by the spin-relaxation. It is well known that the spin-relaxation time decreases as the temperature increases and the magnetic field decreases. However, reduction of the magnetic field strength is unavoidable for detection of recoiled protons, as is discussed in Sec.~\ref{sec:reqpoltargets}. Thus, in order to keep the proton polarization for a sufficiently long time period, it is important to maintain the target at a sufficiently low temperature. 

A beam irradiation test of the polarized target was conducted with a 38-MeV ${\rm ^{12}C}$ beam at the PSI. The polarized target was operated at 0.8~T to avoid the rapid spin relaxation at lower magnetic field. Polarization effects were observed around peaks corresponding to the  $E_{x}=3.50$ ($J^{\pi}=3/2^{-}$) and 3.55~MeV ($J^{\pi}=5/2^{+}$) in ${\rm ^{13}N}$. A rapid decrease in the proton polarization due to irradiation of the ${\rm ^{12}C}$ beam was also observed~\cite{UrregoPhD}.

Further studies to prevent the loss of proton polarization due to the beam irradiation are needed.

\subsection{Polarized ${\rm ^{3}He}$ active targets \label{sec:active3He}}
Recent progress in laser technologies allows spin-polarized ${\rm ^{3}He}$ atoms to be produced at a rate of 10$^{19-20}$ sec$^{-1}$.  Its straightforward application is a polarized gas target stored in a container. However, the container walls are too thick, in many cases, to be used for low-energy experiments. 
For example. in a polarized  ${\rm ^{3}He}$ target used in Ref.~\cite{Uesaka02}, the wall is 100-$\mu$m glass, which corresponds to a range of 4.2~MeV/u ${\rm ^{3}He}$ ion.

One way to overcome the wall problem is to construct an active target with the polarized gas. With the proper selection of materials comprising the electrodes and gas multipliers, the spin-relaxation can be reasonably suppressed.  Interesting applications of this polarized active target include $({\rm ^{3}He},\alpha)$ and $({\rm ^{3}He},{\rm ^{2}He})$ reactions: in these reactions, $\alpha$ and ${\rm ^{2}He}$ are spin zero particles and thus the analyzing power has only weak angular dependences. The effectiveness of the $({\rm ^{3}He},\alpha)$ reaction has been demonstrated at the University of Birmingham, Radial Ridge Cyclotron Facility~\cite{Karban76}.

\section{Summary}

The effects of the spin-dependent interactions in nuclear structure and reactions are overviewed. The strong spin-orbit coupling in nuclei originates from the contributions of the three-nucleon, tensor, and two-nucleon spin-orbit interactions. Recent experimental and theoretical studies indicate that the spin-orbit coupling is subject to substantial changes in nuclei far from the $\beta$ stability line, which has inspired experimental studies of spin-orbit coupling with RI beams and polarized targets.
Among the methods to polarize nuclei developed so far, the triplet-DNP technique is one of the most promising ones for use in RI beam experiments. The first experiments with the triplet-DNP target on the $p$-${\rm ^{6,8}He}$ elastic scattering indicate that the spin-orbit potentials in nuclei with diffused neutron distributions become shallower and more diffuse than the global
systematics of nuclei along the $\beta$-stability line. Future experiments with the $(p,pN)$ quasi-free scattering, proton resonant scattering and transfer reactions will provide a better understanding of the spin-orbit coupling in exotic nuclei. Polarized targets based on other techniques, particularly the frozen-spin operation of DNP targets, active targets of polarized ${\rm ^{3}He}$ gas and gas jet targets for storage ring experiments should extend the polarization studies of exotic nuclei.

%\bibliography{ExoticNuclei}

\begin{thebibliography}{85}

\bibitem{Brack93}
M.~Brack, Reviews of Modern Physics \textbf{65}, 677 (1993)

\bibitem{Mayer49}
M.G. Mayer, Physical Review \textbf{75}, 1969 (1949)

\bibitem{Haxel49}
O.~Haxel, J.H.D. Jensen, H.E. Suess, Physical Review \textbf{75}, 1766 (1949)

\bibitem{Mayer50}
M.G. Mayer, Physical Review \textbf{78}, 16 (1950)

\bibitem{BarschallBrown86}
H.~Barschall, L.~Brown, Foundations of Physics \textbf{16}, 115 (1986)

\bibitem{Chamberlain56}
O.~Chamberlain, E.~Segr\'e, R.D. Tripp, C.~Wiegand, T.~Ypsilantis, Physical
  Review \textbf{102}, 1659 (1956)

\bibitem{Heusinkveld52}
M.~Heusinkveld, G.~Freier, Physical Review \textbf{85}, 80 (1952)

\bibitem{Oxley53}
C.~Oxley, W.~Cartwight, J.~Rouvina, E.~Baskir, D.~Klein, J.~Ring, W.~Skillman,
  Physical Review \textbf{91}, 419 (1953)

\bibitem{Chamberlain57}
O.~Chamberlain, E.~Segr\'e, R.D. Tripp, C.~Wiegand, T.~Ypsilantis, Physical
  Review \textbf{105}, 288 (1957)

\bibitem{Fermi54}
E.~Fermi, Il Nuovo Cimento \textbf{10}, 407 (1954)

\bibitem{Inglis36}
D.R. Inglis, Physical Review \textbf{50}, 783 (1936)

\bibitem{Thomas26}
L.~Thomas, Nature \textbf{117}, 514 (1926)

\bibitem{FujitaMiyazawa57-2}
J.~Fujita, H.~Miyazawa, Progress of Theoretical Physics \textbf{17}, 366 (1957)

\bibitem{FujitaMiyazawa57}
J.~Fujita, H.~Miyazawa, Progress of Theoretical Physics \textbf{17}, 360 (1957)

\bibitem{Terasawa60}
T.~Terasawa, Progress of Theoretical Physics \textbf{23}, 87 (1960)

\bibitem{Machleidt89}
R.~Machleidt, \emph{The Meson Theory of Nuclear Forces and Nuclear Structure}
  (Springer US, 1989), Vol.~19 of \emph{Advances in Nuclear Physics}, book
  section~2, pp. 189--376, ISBN 978-1-4613-9909-4

\bibitem{Scheerbaum76}
R.~R.Scheerbaum, Nuclear Physics A \textbf{257}, 77 (1976)

\bibitem{AndoBando81}
K.~Ando, H.~Bando, Progress of Theoretical Physics \textbf{66}, 227 (1981)

\bibitem{Pieper93}
S.C. Pieper, V.R. Pandharipande, Physical Review Letters \textbf{70}, 2541
  (1993)

\bibitem{Kellogg39}
J.M.B. Kellogg, I.I. Rabi, N.F. Ramsey, J.R. Zacharias, Physical Review
  \textbf{55}, 318 (1939)

\bibitem{Schwinger39}
J.~Schwinger, Physical Review \textbf{55}, 235 (1939)

\bibitem{Yukawa35}
H.~Yukawa, Proc. Physical and Mathematical Society of Japan \textbf{17}, 48
  (1935)

\bibitem{Bethe71}
H.A. Bethe, Annual Review of Nuclear Science \textbf{21}, 93 (1971)

\bibitem{Dobaczewski94}
J.~Dobaczewski, I.~Hamamoto, W.~Nazarewicz, J.A. Sheikh, Phys. Rev. Lett.
  \textbf{72}, 981 (1994)

\bibitem{Lalazissis98}
G.A. Lalazissis, Physics Letters B \textbf{418}, 7 (1998)

\bibitem{Pudliner96}
B.S. Pudliner, A.~Smerzi, J.~Carlson, V.R. Pandharipande, S.C. Pieper, D.G.
  Ravenhall, Phys. Rev. Lett. \textbf{76}, 2416 (1996)

\bibitem{Dobaczewski96}
J.~Dobaczewski, W.~Nazarewicz, T.R. Werner, J.F. Berger, C.R. Chinn,
  J.~Decharg\'e, Physical Review C \textbf{53}, 2809 (1996)

\bibitem{Schiffer04}
J.P. Schiffer, S.J. Freeman, J.A. Caggiano, C.~Deibel, A.~Heinz, C.L. Jiang,
  R.~Lewis, A.~Parikh, P.D. Parker, K.E. Rehm et~al., Physical Review Letters
  \textbf{92}, 162501 (2004)

\bibitem{Otsuka05}
T.~Otsuka, T.~Suzuki, R.~Fujimoto, H.~Grawe, Y.~Akaishi, Phys. Rev. Lett.
  \textbf{95}, 232502 (2005)

\bibitem{Otsuka06}
T.~Otsuka, T.~Matsuo, D.~Abe, Physical Review Letters \textbf{97}, 162501
  (2006)

\bibitem{Noro-ECT2010}
T.~Noro, \emph{{\rm in the ECT* Workshop}} (2010)

\bibitem{Burgunder14}
G.~Burgunder, O.~Sorlin, F.~Nowacki, S.~Giron, F.~Hammache, M.~Moukaddam,
  N.~de~S\'er\'eville, D.~Beaumel, L.~C\`aceres, E.~Cl\'ement et~al., Physical
  Review Letters \textbf{112}, 042502 (2014)

\bibitem{Haeberli67}
W.~Haeberli, Annual Review of Nuclear Science \textbf{17}, 373 (1967)

\bibitem{Happer72}
W.~Happer, Reviews of Modern Physics \textbf{44}, 169 (1972)

\bibitem{Goertz02}
S.~Goertz, W.~Meyer, G.~Reicherz, Progress in Particle and Nuclear Physics
  \textbf{49}, 403 (2002)

\bibitem{Crabb97}
D.G. Crabb, W.~Meyer, Annual Review of Nuclear and Particle Science
  \textbf{47}, 67 (1997)

\bibitem{Rose49}
M.E. Rose, Physical Review \textbf{75}, 213 (1949)

\bibitem{Honig76}
A.~Honig, H.~Mano, Physical Review B \textbf{14}, 1858 (1976)

\bibitem{Clausnitzer56}
H.S. G.~Clausnitzer, R.~Fleischmann, Zeitschrift fur Physik \textbf{144}, 336
  (1956)

\bibitem{Zelenski05}
A.~Zelenski, A.~Bravar, D.~Graham, W.~Haeberli, S.~Kokhanovski, Y.~Makdisi,
  G.~Mahler, A.~Nass, J.~Ritter, T.~Wise et~al., Nuclear Instruments and
  Methods in Physics Research Section A: Accelerators, Spectrometers, Detectors
  and Associated Equipment \textbf{536}, 248 (2005)

\bibitem{Bouchiat60}
M.A. Bouchiat, T.R. Carver, C.M. Varnum, Physical Review Letters \textbf{5},
  373 (1960)

\bibitem{Colegrove63}
F.D. Colegrove, L.D. Schearer, G.K. Walters, Physical Review \textbf{132}, 2561
  (1963)

\bibitem{Albert94}
M.S. Albert, G.D. Cates, B.~Driehuys, W.~Happer, B.~Saam, C.S. Springer,
  A.~Wishnia, Nature \textbf{370}, 199 (1994)

\bibitem{Abragam62}
A.~Abragam, M.~Borghini, P.~Catillon, J.~Coustham, P.~Roubeau, J.~Thirion,
  Physics Letters \textbf{2}, 310 (1962)

\bibitem{Chamberlain63}
O.~Chamberlain, C.~Jeffries, C.~Schultz, G.~Shapiro, L.~Rossum, Physics Letters
  \textbf{7}, 1 (1963)

\bibitem{Uesaka04}
T.~Uesaka, M.~Hatano, T.~Wakui, H.~Sakai, A.~Tamii, Nuclear Instruments and
  Methods in Physics Research Section a-Accelerators Spectrometers Detectors
  and Associated Equipment \textbf{526}, 186 (2004)

\bibitem{ObertelliUesaka11}
A.~Obertelli, T.~Uesaka, European Physical Journal A \textbf{47}, 105 (2011)

\bibitem{Kesteren85}
H.W. van Kesteren, W.T. Wenckebach, J.~Schmidt, Physical Review Letters
  \textbf{55}, 1642 (1985)

\bibitem{Henstra90}
A.~Henstra, T.S. Lin, J.~Schmidt, W.T. Wenckebach, Chemical Physics Letters
  \textbf{165}, 6 (1990)

\bibitem{Iinuma00}
M.~Iinuma, Y.~Takahashi, I.~Shak\'e, M.~Oda, A.~Masaike, T.~Yabuzaki, H.M.
  Shimizu, Physical Review Letters \textbf{84}, 171 (2000)

\bibitem{Wakui05}
T.~Wakui, M.~Hatano, T.~Uesaka, H.~Sakai, A.~Tamii, Nuclear Instruments and
  Methods in Physics Research A \textbf{550}, 521 (2005)

\bibitem{Takeda02}
K.~Takeda, K.~Takegoshi, T.~Terao, Journal of Chemical Physics \textbf{117},
  4940 (2002)

\bibitem{Ghim96}
B.T. Ghim, G.A. Rinard, R.W. Quine, S.S. Eaton, G.R. Eaton, Journal of Magnetic
  Resonance A \textbf{120}, 72 (1996)

\bibitem{Uesaka10}
T.~Uesaka, S.~Sakaguchi, Y.~Iseri, K.~Amos, N.~Aoi, Y.~Hashimoto, E.~Hiyama,
  M.~Ichikawa, Y.~Ichikawa, S.~Ishikawa et~al., Physical Review C \textbf{82},
  021602(R) (2010)

\bibitem{Sakaguchi11}
S.~Sakaguchi, Y.~Iseri, T.~Uesaka, M.~Tanifuji, K.~Amos, N.~Aoi, Y.~Hashimoto,
  E.~Hiyama, M.~Ichikawa, Y.~Ichikawa et~al., Physical Review C \textbf{84},
  024604 (2011)

\bibitem{Sakaguchi13}
S.~Sakaguchi, T.~Uesaka, N.~Aoi, Y.~Ichikawa, K.~Itoh, M.~Itoh, T.~Kawabata,
  T.~Kawahara, Y.~Kondo, H.~Kuboki et~al., Physical Review C \textbf{87},
  021601(R) (2013)

\bibitem{Iinuma95}
M.~Iinuma, Y.~Takahashi, I.~Shak\'e, M.~Oda, A.~Masaike, T.~Yabuzaki, H.M.
  Shimizu, Physics Letters A \textbf{208}, 251 (1995)

\bibitem{Eichhorn14}
T.R. Eichhorn, N.~Niketic, B.~van~den Brandt, U.~Filges, T.~Panzner,
  E.~Rantsiou, W.T. Wenckebach, P.~Hautle, Nuclear Instruments and Methods in
  Physics Research Section A: Accelerators, Spectrometers, Detectors and
  Associated Equipment \textbf{754}, 10 (2014)

\bibitem{Uesaka-EURISOL}
T.~Uesaka, K.~Tateishi, S.~Sakaguchi, E.~Milman, S.~Chebotaryov, T.~Kawahara,
  T.~Tang, T.~Wakui, T.~Tshukihana, Y.~Urata et~al., \emph{{\rm Proceedings of
  the 5th EURISOL Topical Meeting}} (2014)

\bibitem{Tateishi14}
K.~Tateishi, M.~Negoro, S.~Nishida, A.~Kagawa, Y.~Morita, M.~Kitagawa,
  Proceedings of the National Academy of Sciences \textbf{111}, 7527 (2014)

\bibitem{Kubo92}
T.~Kubo, M.~Ishihara, N.~Inabe, H.~Kumagai, I.~Tanihata, K.~Yoshida,
  T.~Nakamura, H.~Okuno, S.~Shimoura, K.~Asahi, Nuclear Instruments and Methods
  in Physics Research B \textbf{70}, 309 (1992)

\bibitem{Korsheninnikov93}
A.~Korsheninnikov, K.~Yoshida, D.~Aleksandrov, N.~Aoi, Y.~Doki, N.~Inabe,
  M.~Fujimaki, T.~Kobayashi, H.~Kumagai, C.B. Moon et~al., Physics Letters B
  \textbf{316}, 38 (1993)

\bibitem{Korsheninnikov97}
A.~Korsheninnikov, E.Y. Nikolski, C.A. Bertulani, S.~Fukuda, T.~Kobayashi, E.A.
  Kuzmin, S.~Momota, B.G. Novatskii, A.A. Ogloblin, A.~Ozawa et~al., Nuclear
  Physics A \textbf{617}, 45 (1997)

\bibitem{KoningDelaroche03}
A.J. Koning, J.P. Delaroche, Nuclear Physics A \textbf{713}, 231 (2003)

\bibitem{Varner91}
R.L. Varner, W.J. Thompson, T.L. McAbee, E.J. Ludwig, T.B. Clegg, Physics
  Reports \textbf{201}, 57 (1991)

\bibitem{JacobMaris66}
G.~Jacob, T.A.J. Maris, Rev. Mod. Phys. \textbf{38}, 121 (1966)

\bibitem{JacobMaris73}
G.~Jacob, T.A.J. Maris, Rev. Mod. Phys. \textbf{45}, 6 (1973)

\bibitem{Kitching80}
P.~Kitching, C.A. Miller, W.C. Olsen, D.A. Hutcheon, W.J. McDonald, A.W. Stetz,
  Nuclear Physics A \textbf{340}, 423 (1980)

\bibitem{Jacob73}
G.~Jacob, T.A.J. Maris, C.~Schneider, M.R. Teodoro, Physics Letters
  \textbf{45B}, 181 (1973)

\bibitem{Jacob76}
G.~Jacob, T.A.J. Maris, C.~Schneider, M.R. Teodoro, Nuclear Physics A
  \textbf{257}, 517 (1976)

\bibitem{Kitching76}
P.~Kitching, C.A. Miller, D.A. Hutcheon, A.N. James, W.J. McDonald, J.M.
  Cameron, W.C. Olsen, G.~Roy, Physical Review Letters \textbf{37}, 1600 (1976)

\bibitem{Panin16}
V.~Panin, J.T. Taylor, S.~Paschalis, F.~Wamers, Y.~Aksyutina, H.~Alvarez-Pol,
  T.~Aumann, C.A. Bertulani, K.~Boretzky, C.~Caesar et~al., Physics Letters B
  \textbf{753}, 204 (2016)

\bibitem{Kawase15}
S.~Kawase, T.~Tang, T.~Uesaka, D.~Beamel, M.~Dozono, T.~Fujii, T.~Fukunaga,
  N.~Fukuda, A.~Galindo-Uribarri, S.~Hwang et~al., JPS Conference Proceedings
  \textbf{6}, 020003 (2015)

\bibitem{Tang15}
T.~Tang, S.~Kawase, T.~Uesaka, D.~Beamel, M.~Dozono, T.~Fujii, T.~Fukunaga,
  N.~Fukuda, A.~Galindo-Uribarri, S.~Hwang et~al., JPS Conference Proceedings
  \textbf{6}, 030077 (2015)

\bibitem{Goldberg93}
V.~Goldberg, A.~Pakhomov, Physics of Atomic Nuclei \textbf{56}, 1167 (1993)

\bibitem{Charity16}
R.J. Charity, European Physical Journal Plus \textbf{131}, 63 (2016)

\bibitem{Teranishi13}
T.~Teranishi, S.~Sakaguchi, T.~Uesaka, H.~Yamaguchi, S.~Kubono, T.~Hashimoto,
  S.~Hayakawa, Y.~Kurihara, D.N. Bihn, D.~Kahl et~al., AIP Conference
  Proceedings \textbf{1525}, 552 (2013)

\bibitem{YuleHaeberli67}
T.J. Yule, W.~Haeberli, Physical Review Letters \textbf{19}, 756 (1967)

\bibitem{YuleHaeberli68}
T.J. Yule, W.~Haeberli, Nuclear Physics A \textbf{117}, 1 (1968)

\bibitem{UrregoBlanco05}
J.P. Urrego-Blanco, B.~van~den Brandt, E.I. Bunyatova, A.~Galindo-Uribarri,
  P.~Hautle, J.A. Konter, Nuclear Instruments and Methods in Physics Research
  Section B: Beam Interactions with Materials and Atoms \textbf{241}, 1001
  (2005)

\bibitem{UrregoBlanco07}
J.P. Urrego-Blanco, C.R. Bingham, B.~van~den Brandt, A.~Galindo-Uribarri,
  J.~G\'omez~del Campo, P.~Hautle, J.A. Konter, E.~Padilla-Rodal, P.A.
  Schmelzbach, Nuclear Instruments and Methods in Physics Research Section B:
  Beam Interactions with Materials and Atoms \textbf{261}, 1112 (2007)

\bibitem{Bunyatova04}
E.I. Bunyatova, Nuclear Instruments and Methods in Physics Research Section A:
  Accelerators, Spectrometers, Detectors and Associated Equipment \textbf{526},
  22 (2004)

\bibitem{UrregoPhD}
J.P. Urrego-Blanco, Ph.d thesis (2009)

\bibitem{Uesaka02}
T.~Uesaka, J.~Nishikawa, H.~Okamura, K.~Suda, H.~Sakai, A.~Tamii, K.~Sekiguchi,
  K.~Yako, S.~Sakoda, H.~Kato et~al., Physics Letters B \textbf{533}, 1 (2002)

\bibitem{Karban76}
O.~Karban, K.~Basak, J.~England, G.~Morrison, J.~Nelson, S.~Roman, G.~Shute,
  Nuclear Physics A \textbf{269}, 312 (1976)

\end{thebibliography}
%\bibliographystyle{epj}

\end{document}